\documentclass[twocolumn,showpacs,preprintnumbers,amsmath,amssymb,floatfix,prd,nofootinbib]{revtex4-1}
\usepackage{epsfig}
\usepackage{dcolumn}
\usepackage{bbold} %for unit matrix symbol \mathbb{1}
\usepackage{wasysym} %\permil
\usepackage{hyperref}
\usepackage{color,soul}
\def \be  {\begin{equation}}
\def \ee  {\end{equation}}
\def \ba  {\begin{eqnarray}}
\def \ea  {\end{eqnarray}}
\def \baa {\begin{eqnarray*}}
\def \eaa {\end{eqnarray*}}
\def \nn {\nonumber}
\def \eps {\varepsilon}
\def \nbar {\bar{N}}
\def \LL {\text{LL}}
\def \NLL {\text{NLL}}
\def \NNLL {\text{NNLL}}
\def \bs {\boldsymbol}

\newcommand\f{\frac}
\begin{document}
\title{Fragmentation Functions Beyond Fixed Order Accuracy}
\begin{flushright}
LA-UR-16-27763 \\
\end{flushright}
\author{Daniele P.\ Anderle}
%\email{daniele-paolo.anderle@uni-tuebingen.de}
\affiliation{Institute for Theoretical Physics, University of T\"ubingen, Auf der Morgenstelle 
14, 72076 T\"ubingen, Germany}
\affiliation{School of Physics and Astronomy, The University of Manchester, Manchester, M13 9PL, U.K.}
\author{Tom Kaufmann}
%\email{tom.kaufmann@uni-tuebingen.de}
\affiliation{Institute for Theoretical Physics, University of T\"ubingen, Auf der Morgenstelle 
14, 72076 T\"ubingen, Germany}
\author{Felix Ringer}
%\email{f.ringer@lanl.gov}
\affiliation{Theoretical Division, MS B283, Los Alamos National Laboratory, Los Alamos, NM 87545, USA}
\author{Marco Stratmann}
%\email{marco.stratmann@uni-tuebingen.de}
\affiliation{Institute for Theoretical Physics, University of T\"ubingen, Auf der Morgenstelle 
14, 72076 T\"ubingen, Germany}

\begin{abstract}
We give a detailed account of the phenomenology of all-order
resummations of logarithmically enhanced contributions at small
momentum fraction of the observed hadron in semi-inclusive electron-positron
annihilation and the time-like scale evolution of parton-to-hadron fragmentation functions.
The formalism to perform resummations in Mellin moment space 
is briefly reviewed, and all relevant expressions up to next-to-next-to-leading
logarithmic order are derived, including their explicit dependence on
the factorization and renormalization scales.
We discuss the details pertinent to a proper 
numerical implementation of the resummed results comprising
an iterative solution to the time-like evolution equations, the
matching to known fixed-order expressions, and the choice of
the contour in the Mellin inverse transformation.
First extractions of parton-to-pion fragmentation functions from 
semi-inclusive annihilation data are performed at different 
logarithmic orders of the resummations in order to estimate their
phenomenological relevance. 
To this end, we compare our results to corresponding
fits up to fixed, next-to-next-to-leading order accuracy 
and study the residual dependence on the factorization 
scale in each case.
\end{abstract}

\pacs{13.87.Fh, 13.85.Ni, 12.38.Bx}

\maketitle

%%%%%%%%%%%%%%%%%%%%%%%%%%%%%%%%%%%%%
\section{Introduction and Motivation}
%%%%%%%%%%%%%%%%%%%%%%%%%%%%%%%%%%%%%
%
Fragmentation functions (FFs) $D_i^h(z,Q^2)$ are an integral part of the theoretical framework 
describing hard-scattering processes with an observed hadron in the final-state in perturbative 
QCD (pQCD) \cite{ref:fact}. They parametrize in a process-independent way
the non-perturbative transition of a parton with a particular flavor $i$ into a hadron
of type $h$ and depend on the fraction $z$ of the parton's longitudinal momentum taken by the hadron and a
large scale $Q$ inherent to the process under consideration \cite{ref:collins-soper}.
The prime example is single-inclusive electron-positron annihilation (SIA), 
$e^-e^+\to h X$, at some center-of-mass system (c.m.s.) energy $\sqrt{S}=Q$,
where $X$ is some unidentified hadronic remnant. 

Precise data on SIA \cite{ref:belledata,ref:babardata,ref:tpcdata,ref:slddata,ref:alephdata,ref:delphidata,ref:opaldata}, 
available at different $\sqrt{S}$, ranging from about $10\,\mathrm{GeV}$ up to
the mass $M_Z$ of the $Z$ boson, reveal important experimental information on FFs that is routinely
used in theoretical extractions, i.e., fits of FFs \cite{ref:dss,ref:dss2,Epele:2012vg,ref:dssnew,Anderle:2015lqa,ref:other-ffs}.
Processes other than SIA are required, however, to gather the information needed to fully disentangle
all the different FFs $D_i^h$ for $i=u,\bar{u},d,\bar{d},\ldots$ quark and antiquark flavors and the gluon.
Specifically, data on semi-inclusive deep-inelastic scattering (SIDIS), $e^{\pm}p\to hX$, and the
single-inclusive, high transverse momentum ($p_T$) production of hadrons in proton-proton collisions,
$pp\to hX$, are utilized, which turn extractions of FFs into global QCD analyses \cite{ref:dss,ref:dss2,Epele:2012vg,ref:dssnew}. 
Most recently, a proper theoretical framework in terms of FFs has been developed for a novel
class of processes, where a hadron is observed inside a jet \cite{ref:jethadron_th}. 
It is expected that corresponding data \cite{ref:jethadron_exp}
will soon be included in global analyses, where they will provide additional constraints on,
in particular, the gluon-to-hadron FF.

The ever increasing precision of all these probes sensitive to the hadronization of (anti-)quarks and gluons
has to be matched by more and more refined theoretical calculations. 
One way of advancing QCD calculations is the computation of higher order corrections in the strong coupling
$\alpha_s$. Here, next-to-leading order (NLO) results are available throughout for all ingredients
needed for a global QCD analysis of FFs as outlined above. Specifically, they comprise 
the partonic hard scattering cross sections for inclusive hadron production in SIA \cite{ref:sia-nlo,ref:nason}, 
SIDIS \cite{ref:sia-nlo,ref:nason,Graudenz:1994dq,deFlorian:1997zj}, and $pp$ collisions \cite{ref:pp-nlo} 
and the evolution kernels or time-like parton-to-parton splitting functions 
$P_{ij}^{T}$ \cite{ref:splitting-nlo,ref:sia-neerven-long-as2,Stratmann:1996hn,ref:nnlo-sia}, 
which govern the scale $Q$ dependence of the FFs through a set of integro-differential evolution equations \cite{ref:DGLAP}.
Such type of NLO global analyses of FFs represent the current state-of-the-art in this field.
For instance, a recent extraction of parton-to-pion FFs $D_i^{\pi}$ at NLO accuracy can be
found in Ref.~\cite{ref:dssnew}. A special role in this context plays SIA, where fits of FFs can be carried out already
at the next-to-next-leading order (NNLO) level thanks to the available 
SIA coefficient functions \cite{ref:sia-neerven-long-as2,ref:nnlo-sia,ref:nnlo-sia+mellin1, ref:nnlo-sia-mellin2}
and kernels $P_{ij}^{T}$ at NNLO \cite{ref:nnlo-kernel}. This has not yet been achieved in the case of hadron production 
in SIDIS or in $pp$ collisions. A first determination of parton-to-pion FFs from SIA data at NNLO accuracy 
has been performed  recently in \cite{Anderle:2015lqa}.

Another important avenue for systematic improvements in the theoretical analysis of data sensitive to FFs, 
which we pursue in this paper, concerns large logarithms present in each fixed order 
of the perturbative series in $\alpha_s$
for both the evolution kernels $P_{ij}^{T}$ and the process-dependent hard scattering coefficient functions.
In this paper we will deal with logarithms that become large in the limit of small momentum fractions $z$
and, in this way, can spoil the convergence of the expansion in $\alpha_s$ even when the coupling is very 
small. 
As we shall see, two additional powers of $\log^{2k}(z)$ can arise in each fixed order $\alpha_s^k$, which is 
numerically considerably more severe than in the space-like case relevant to deep-inelastic scattering (DIS)
and the scale evolution of parton density functions (PDFs) and completely destabilizes the
behavior of cross sections and FFs in the small-$z$ regime.

To mitigate the singular small-$z$ behavior imprinted by these logarithms, one needs to resum them
to all orders in perturbation theory, a well-known procedure \cite{ref:mueller}.
Knowledge of the fixed-order results up to $\mathrm{N}^m\mathrm{LO}$ determines, in principle,
the first $m+1$ ``towers'' of logarithms to all orders.
Hence, thanks to the available NNLO results, small-$z$ resummations have been pushed up to the first
three towers of logarithms for SIA and the time-like splitting functions $P_{ij}^{T}$ recently,
which is termed the next-to-next-to-leading logarithmic (NNLL) approximation \cite{Vogt:2011jv, Kom:2012hd}.
Based on general considerations on the structure of all-order mass factorization, as proposed and
utilized in Ref.~\cite{Vogt:2011jv, Kom:2012hd},
we re-derive the resummed coefficient functions for SIA and the evolution kernels $P_{ij}^{T}$ and
compare them to the results available in the literature. Next, we shall extend these expressions by restoring
their dependence on the factorization and renormalization scales $\mu_F$ and $\mu_R$, respectively,
which will allow us to estimate the theoretical uncertainties related to the choice of $\mu_F/Q$.
It is expected that the scale ambiguity will shrink the more higher order corrections are included.
We note that large logarithms also appear in the limit $z\to 1$. Their phenomenological implications
have been addressed in the case of SIA in Ref.~\cite{ref:vogt_large_x, Anderle:2012rq}, and we shall not consider them in the present study
focussing mainly on the small-$z$ regime.

Resummations are most conveniently carried out in Mellin-$N$ moment space, which also gives the 
best analytical insight into the solution of the coupled, matrix-valued scale evolution equations obeyed
by the quark singlet and gluon FFs. We shall discuss in some detail how we define a solution
to these evolution equations beyond the fixed-order approximation, i.e., based on resumed
kernels $P_{ij}^{T}$. We also explain how we match the resummed small-$z$ expressions to a given
fixed-order result defined for all $z$, thereby avoiding any double-counting of logarithms and also 
maintaining the validity of the momentum sum rule.
We shall also address in our discussions 
the proper numerical implementation of the resummed expressions in Mellin $N$ space,
in particular, the structure of singularities and the choice of the integration contour for
the inverse Mellin transformation back to the physical $z$ space. 
Already at fixed, NNLO accuracy this is known to be a non-trivial issue \cite{Anderle:2015lqa}.

After all these technical preparations, we will present some phenomenological applications.
So far, resummations in the context of FFs have been, to the best of our knowlegde, 
exclusively studied for the $N=1$ moment, the $z$ integrated hadron multiplicities, in particular, their
scale evolution and the shift of the peak of the multiplicity distribution with energy \cite{ref:mueller, ref:n1moment_pheno}. 
At fixed order, multiplicities are ill-defined due to the singularities induced by the small-$z$ behavior. 
In the ``modified leading logarithmic
approximation'' (MLLA) and beyond, i.e., upon including resummed expressions, 
these singularities are lifted, and one finds 
a rather satisfactory agreement with data, which can be used to determine, e.g., 
the strong coupling $\alpha_s$ in SIA \cite{ref:n1moment_pheno}.
We plan to revisit the phenomenology of $N=1$ multiplicities in a separate publication elsewhere.
In this paper, we will apply resummations in the entire $z$ range, i.e., for the first time, 
we extract FFs from SIA data with identified pions up to NNLO+NNLL accuracy, including a proper matching procedure. 
We shall investigate the phenomenological relevance of small-$z$ resummations in achieving the best possible 
description of the SIA data.
This will be done by comparing the outcome of a series of fits to data both at fixed order accuracy
and by including up to three towers of small-$z$ logarithms. We also compare the so obtained
quark singlet and gluon FFs and estimate the residual theoretical uncertainty due to the choice
of $\mu_F/Q$ in each case.
An important phenomenological question that arises in this context is how low in $z$ one can push the 
theoretical framework outlined above before neglected kinematic hadron mass corrections become relevant.
Hadron mass effects in SIA have been investigated to some extent in \cite{ref:hmc+res} but there is no systematic way to
properly include them in a general process \cite{Christova:2016hgd}, i.e., ultimately in a global analysis of FFs.
Therefore, one needs to determine a lower value of $z$, largely on kinematical considerations, below
which fits of FFs make no sense. We will discuss this issue as well in the phenomenological section of the
paper.
In general, it turns out, that in the range of $z$ where SIA data are available
and where the framework can be applied, a 
fit at fixed, NNLO accuracy already captures most of the relevant small-$z$ behavior needed
to arrive at a successful description of the data, and resummations add only very little in a fit.
 
The remainder of the paper is organized as follows: Section \ref{sec:nnlo} comprises
all relevant technical aspects. We start by briefly reviewing the fixed order results
for semi-inclusive annihilation and catalogue the systematics of the small-$z$ logarithms 
that appear in each order of perturbation theory. 
Next, we show how these logarithms can be resummed to all orders
and compare to existing results in the literature.
In Sec.~\ref{subsec:scales} we provide the expressions containing logarithms of the
factorization and renormalization scales to estimate the remaining
theoretical uncertainties after resummation. The solution of the time-like evolution
equations with resummed splitting functions in Mellin moment space
is discussed in Sec.~\ref{subsec:evol}. Peculiarities important for a proper numerical
implementation of the resummed expressions in $N$-space are raised in Sec.~\ref{subsec:num}.
In the second part of the paper we discuss the phenomenological implications 
of small-$z$ resummations for the extraction
of fragmentation functions from data. In Sec.~\ref{subsec:fit} we present and discuss
various fits to semi-inclusive annihilation data at different fixed-orders in perturbation theory and
levels of small-$z$ resummations. Finally, in Sec.~\ref{subsec:appl} we study the residual
scale dependence with and without resummations of small-$z$ logarithms.
We conclude in Sec.~\ref{sec:conclusions}.
 
%%%%%%%%%%%%%%%%%%%%%%%%%%%%%%%%%%%%%
\section{Small-$z$ Resummation for Semi-inclusive $\mathbf{e^+e^-}$ annihilation \label{sec:nnlo}}
%%%%%%%%%%%%%%%%%%%%%%%%%%%%%%%%%%%%%
%
This section covers all the relevant technical aspects of small-$z$ resummations in SIA: 
in Sec.~\ref{subsec:xsec} we briefly recall
the calculation of the cross section for SIA up to NNLO accuracy, introduce FFs, and the
Mellin transformation. In addition, we sketch the systematics of the small-$z$ enhanced logarithmic contributions 
that appear in both the coefficient functions for SIA and in the time-like evolution kernels
in each order of perturbation theory. The resummation of these logarithms up to NNLL accuracy
is concisely reviewed in Sec.~\ref{subsec:resum}, where we also compare our results to those available in
the literature. In Sec.~\ref{subsec:scales}, we extend the currently available
resummed expressions for the SIA cross section
by re-introducing their dependence on the scales $\mu_F$ and $\mu_R$, which is vital for a discussion of theoretical uncertainties later on in the phenomenological section of the paper. 
In Sec.~\ref{subsec:evol} we explain in some detail how the resummed kernels are used in solving the time-like
evoltion equations in Mellin $N$ space, and numerical peculiarities, in particular, those associated with
the Mellin inverse transformation are covered in Sec.~\ref{subsec:num}
%
%%%%%%%%%%%%%%%%%%%%%%%%%%%%%%%%%%%%%
\subsection{Fixed order SIA, fragmentation functions, and the systematics of small-$z$ logarithms \label{subsec:xsec}}
%%%%%%%%%%%%%%%%%%%%%%%%%%%%%%%%%%%%%
%
We consider the SIA process $e^+e^-\to\gamma/Z\to h X$, 
more specifically, cross sections defined as
\be\label{eq:TL}
\frac{d\sigma^h}{dz} = \sum_{k=T,L} \f{d\sigma^h_k}{dz}\,.
\ee
The parity-violating interference term of vector and axial-vector contributions,
usually called ``asymmetric'' ($A$), is not present in (\ref{eq:TL}) as we have already integrated 
over the scattering angle $\theta$; see, e.g.\ \cite{ref:nason}.
Hence, only the transverse ($T$) and the longitudinal ($L$) parts
remain and will be considered in what follows. Furthermore, we have introduced the scaling variable
\be
\label{eq:zdef}
z\equiv \frac{2P_h\cdot q}{Q^2}\stackrel{\mathrm{c.m.s.}}{=} \frac{2E_h}{Q}\,, 
\ee
where $P_h$ and $q$ are the four momenta of the observed hadron and time-like $\gamma/Z$ boson, respectively.
Moreover, $Q^2 \equiv q^2 = S$. As indicated in Eq.~(\ref{eq:zdef}), $z$ reduces to the hadron's energy fraction 
in the c.m.s.\ and is often also labeled as $x_E$ \cite{ref:nason}. 
Note, that experimental data are usually given in terms of hadron multiplicity distributions,
which are equivalent to the cross sections as defined in Eq.~\eqref{eq:TL} normalized by the
total hadronic cross section $\sigma_{\mathrm{tot}}$ \cite{ref:nnlo-sia,ref:nnloint}.

The transverse and longitudinal cross sections in Eq.~\eqref{eq:TL} may be written in a factorized form as \cite{ref:nnlo-sia,ref:nnlo-sia-mellin2}
\ba\label{eq:nnlostructure}
\f{d\sigma^h_k}{dz} = \sigma_{\mathrm{tot}}^{(0)} && \left[D_\mathrm{S}^h(z,\mu^2)\otimes 
                              \mathbb{C}_{k,q}^\mathrm{S}\left(z,\f{Q^2}{\mu^2}\right) \right. \nn \\
&& \!\!\!\!\!\! + \left.  D_g^h\left(z,\mu^2\right)\otimes 
                            \mathbb{C}_{k,g}^\mathrm{S}\left(z,\f{Q^2}{\mu^2}\right) \right] \nn \\
&& \!\!\!\!\!\! + \,\sum_q \sigma_q^{(0)}\,D_{\mathrm{NS},q}^h(z,\mu^2) \otimes  
                            \mathbb{C}_{k,q}^\mathrm{NS}\left(z,\f{Q^2}{\mu^2}\right) . 
\ea
For simplicity, we have chosen the factorization and renormalization scales equal, $\mu_R=\mu_F\equiv \mu$, and 
$\sigma_q^{(0)} = 3 \sigma_0 \hat{e}_q^2$ 
is the total quark production cross section for
a given flavor $q$ at leading order (LO). $\sigma_0= 4\pi\alpha^2/(3Q^2)$ denotes
the lowest order QED cross section for the process $e^+e^-\to \mu^+ \mu^-$ with $\alpha$ the electromagnetic coupling. 
The electroweak quark charges $\hat{e}_q$ can be found, e.g., in Ref.~\cite{ref:nnlo-sia}.
We also defined $\sigma_{\mathrm{tot}}^{(0)}=\sum_q\sigma_q^{(0)}$.
The symbol $\otimes$ denotes the standard convolution integral which is given by
\be\label{eq:convolution}
f(z)\otimes g(z)\equiv \int_0^1dx\int_0^1dy \, f(x)\, g(y)\,\delta(z-xy)\, .
\ee
With this notation, the transverse and longitudinal cross sections are related to the usual longitudinal and transverse structure functions~\cite{ref:sia-neerven-long-as2} according to
\ba\label{eq:FTFL}
F_k &\equiv& \frac{1}{3\sigma_0} \f{d\sigma^h_k}{dz} \nn \\
&=& \left(\sum_q \hat{e}_q^2 \right)
\left[D_\mathrm{S}^h(z,\mu^2)\otimes 
                              \mathbb{C}_{k,q}^\mathrm{S}\left(z,\f{Q^2}{\mu^2}\right) \right. \nn \\
&& ~~~~~~ + \left.  D_g^h\left(z,\mu^2\right)\otimes 
                            \mathbb{C}_{k,g}^\mathrm{S}\left(z,\f{Q^2}{\mu^2}\right) \right] \nn \\
&& ~~~~~~ + \,\sum_q \hat{e}_q^2\,D_{\mathrm{NS},q}^h(z,\mu^2) \otimes  
                            \mathbb{C}_{k,q}^\mathrm{NS}\left(z,\f{Q^2}{\mu^2}\right) \nn \\
&=& \sum_{l=q,\bar{q},g} D_l^h(z,\mu^2)\otimes 
                              \mathbb{C}_{k,l}\left(z,\f{Q^2}{\mu^2}\right)\,.
\ea
As usual, the factorized structure of Eq.~\eqref{eq:nnlostructure} holds in the presence of a hard scale, i.e., 
of ${\cal{O}}$(few GeV), and up to corrections that 
are suppressed by inverse powers of the hard scale. SIA is a one-scale process, and the hard scale should be
chosen to be of ${\cal{O}}(Q)$.
The power corrections for SIA are much less well understood than in DIS, perhaps due to the lack of an
operator product expansion in the time-like case. One source, which we will get back to later on, is 
of purely kinematic origin. Instead of the energy fraction $z$, SIA data are often given in terms of 
the hadron's three-momentum fraction in the c.m.s., $x_p=2p/Q$, which leads to $1/Q^2$ corrections 
when converted back to proper scaling variable: $x_p=z-2m_h^2/(z Q^2) + {\cal{O}}(1/Q^4)$ \cite{ref:nason}. 
$m_h$ is the produced hadron's mass and is neglected in the factorized formalism outlined above.
Other sources of power corrections arise in the non-perturbative formation of a hadron from
quarks or gluons and are expected to behave like $1/Q$ from model estimates \cite{ref:nason}.

The dependence of the FFs on the factorization scale $\mu$ may be calculated in pQCD
and is described by the $2N_f+1$ coupled integro-differential evolution equations \cite{ref:DGLAP} with $N_f$ being
the number of active quark flavors.
It is common to define certain linear combinations of quark and antiquark FFs
that appear in SIA. The
quark singlet ($\mathrm{S}$) and nonsinglet ($\mathrm{NS}$) FFs in Eq.~\eqref{eq:nnlostructure}
are given by
\be 
\label{eq:sing}
D_{\mathrm{S}}^h(z,\mu^2) =  \f{1}{N_f}\sum_q \left[ D_q^h(z,\mu^2) + D_{\bar q}^h(z,\mu^2) \right]
\ee
and
\be
\label{eq:nsing}
D_{\mathrm{NS},q}^h(z,\mu^2)  =  D_q^h(z,\mu^2) + D_{\bar q}^h(z,\mu^2) - D_{\mathrm{S}}^h(z,\mu^2)
\ee
respectively. The corresponding coefficient functions $i=\mathrm{S, NS}$ in~(\ref{eq:nnlostructure}) can be calculated as a perturbative series in $a_s\equiv\alpha_s/4 \pi$,
\be \label{eq:coeffexp}
\mathbb{C}_{k,l}^{i} = \mathbb{C}^{i,(0)}_{k,l}  + a_s\,\mathbb{C}^{i,(1)}_{k,l} + a_s^2\, \mathbb{C}^{i,(2)}_{k,l} + \mathcal{O}(a_s^3)  \, ,
\ee
where we have suppressed the arguments $(z,Q^2/\mu^2)$. 
Expressions for the $\mathbb{C}_{k,l}^i$ are available 
up to ${\cal O}(a_s^2)$ in Refs.~\cite{ref:sia-neerven-long-as2,ref:nnlo-sia,ref:nnlo-sia+mellin1}, 
which is NNLO for the transverse coefficient functions but formally only next-to-leading accuracy (NLO) accuracy 
for the longitudinal coefficient functions as the latter start to be non-zero at ${\cal O}(a_s)$.

The fixed order results of the coefficient functions contain logarithms that become 
large for $z\to 1$ (large-$z$ regime) and $z\to 0$ (small-$z$ regime). 
Such large logarithms can potentially spoil the convergence of the
perturbative expansion even for $a_s\ll 1$ and, hence, need to be taken into account to all orders in the strong coupling. 
The resummation of large-$z$ logarithms in SIA has been addressed, for instance, in 
Refs.~\cite{Anderle:2012rq,ref:vogt_large_x}. 
The main focus of this paper is on the so far very little explored small-$z$ regime and its 
phenomenology. In contrast to the space-like DIS process with its single logarithmic enhancement,
one finds a double logarithmic enhancement for the time-like SIA; see, e.g., 
\cite{Blumlein:1997em} and references therein. For example, for the gluon sector in 
Eq.~\eqref{eq:nnlostructure} one finds
\ba\label{eq:logC}
\mathbb{C}_{T,g}^{S,(k)} &\propto& a_s^k \frac{1}{z} \log^{2k-1-a}(z)\,,\nn \\ 
\mathbb{C}_{L,g}^{S,(k)} &\propto& a_s^k \frac{1}{z} \log^{2k-2-a}(z)\,,
\ea
where $a=0,1,$ and $2$ corresponds to the leading logarithmic (LL),
next-to-leading logarithmic (NLL), and NNLL contribution, respectively. 

Furthermore, the same logarithmic behavior at small-$z$ is found for the 
time-like splitting functions that govern the scale evolution of the FFs. 
For example, for the gluon-to-gluon and the quark-to-gluon splitting function, one finds
\be\label{eq:logP}
P_{gi}^{T,(k)} \propto a_s^{(k+1)} \frac{1}{z} \log^{2k-a}(z)\,,
\ee
where $i=q,g$, and $k$ denotes the perturbative order starting from $k=0$, i.e., LO. 
In order to obtain a reliable prediction from perturbative QCD in the small-$z$ regime, 
these large logarithmic contributions, both in the coefficient functions and in the splitting functions, 
need to be resummed to all orders. 
The resulting expressions are available in the literature up to NNLL accuracy~\cite{Vogt:2011jv,Kom:2012hd}
and we will re-derive them in the next subsection. 
Traditionally, and most conveniently, these calculations are carried out in the complex Mellin transform space. 
In general, the Mellin integral transform $f(N)$ of a function $f(z)$ is defined by
\begin{equation}
\label{eq:mellin}
f(N)=\int_0^1dz\,z^{N-1}f(z) \equiv \mathcal{M}[f(z)] \, .
\end{equation}
Hence, the Mellin transform of the small-$z$ logarithms given in Eqs.~\eqref{eq:logC} and \eqref{eq:logP} 
reads
\be
\mathcal{M}\left[\frac{\log^{2k-1}(z)}{z} \right] = (-1)^k \frac{(2k-1)!}{\nbar^{2k}}\,,
\ee
where $\nbar \equiv N-1$, i.e., they give rise to singularities at $N=1$ in Mellin space. 

The structure of the $1/\nbar$ divergences for all quantities relevant to a theoretical
analysis of SIA up to NNLL accuracy is summarized
schematically in Tables~\ref{tab:Ndependence_coefs} and \ref{tab:Ndependence_splittings}. 
\begin{table}[th!]
\caption{\label{tab:Ndependence_coefs} The explicit $1/\nbar$ dependence of the coefficient functions 
\mbox{$\mathbb{C}_{k,l}^{S} = \sum_n a_s^n \mathbb{C}_{k,l}^{S,(n)}$}
at any given fixed order $n$ of the perturbative expansion at the LL, NLL, and NNLL approximation.
These generic structures are valid starting from $n=1$ or $n=2$ 
as indicated in the bottom row of the table. 
For smaller values of $n$, the correct $1/\nbar$ dependence must be extracted from the 
fixed order results; see text. 
Also, note that the entry for $\mathbb{C}_{L,g}^{S,(n)}$ at NNLL is obtained by 
$\mathcal{AC}$ relations; see text.}
\begin{ruledtabular}
\begin{tabular}{lcccc}
 & $\mathbb{C}_{T,g}^{S,(n)}$ & $\mathbb{C}_{T,q}^{S,(n)}$ & $\mathbb{C}_{L,g}^{S,(n)}$  &  $\mathbb{C}_{L,q}^{S,(n)}$ \\[1mm]\hline
LL \rule[0mm]{0mm}{4mm}  & $\nbar^{-2n}$ & -- & $\nbar^{1-2n}$ & -- \\[1mm]
NLL & $\nbar^{1-2n}$ & $\nbar^{1-2n}$ & $\nbar^{2-2n}$ & $\nbar^{2-2n}$ \\[1mm]
NNLL & $\nbar^{2-2n}$ & $\nbar^{2-2n}$ & $\nbar^{3-2n}$ & $\nbar^{3-2n}$ \\[1mm] \hline
\rule[0mm]{0mm}{4mm} & $n\ge1$ & $n\ge2$ &$n\ge1$& $n\ge2$\\
\end{tabular}
\end{ruledtabular}
\end{table}
\begin{table}[th!]
\caption{\label{tab:Ndependence_splittings} Same as Tab.~\ref{tab:Ndependence_coefs} but for the splitting functions \mbox{$P_{ij}^{T} = \sum_n a_s^{n+1} P_{ij}^{T,(n)}$}.}
\begin{ruledtabular}
\begin{tabular}{lcccc}
 & $P_{gg}^{T,(n)}$ & $P_{gq}^{T,(n)}$ & $P_{qq}^{T,(n)}$  &  $P_{qg}^{T,(n)}$ \\[1mm]\hline
LL \rule[0mm]{0mm}{4mm}  & $\nbar^{-1-2n}$ & $\nbar^{-1-2n}$ & -- & -- \\[1mm]
NLL & $\nbar^{-2n}$ & $\nbar^{-2n}$ & $\nbar^{-2n}$ & $\nbar^{-2n}$ \\[1mm]
NNLL & $\nbar^{1-2n}$ & $\nbar^{1-2n}$ & $\nbar^{1-2n}$ & $\nbar^{1-2n}$ \\[1mm] \hline
\rule[0mm]{0mm}{4mm} & $n\ge0$ & $n\ge0$ &$n\ge2$& $n\ge2$\\
\end{tabular}
\end{ruledtabular}
\end{table}
Note that no LL contributions appear in
the quark sector, neither for the splitting nor for the coefficient functions. 
Moreover, the LO and NLO small-$z$ contributions to $\mathbb{C}_{T/L,q}^{S}$, $P^T_{qq}$, and $P^T_{qg}$ 
are not contained in the generic structure summarized in Tables~\ref{tab:Ndependence_coefs} and
\ref{tab:Ndependence_splittings}. 
Instead, these terms have to be extracted directly from the respective fixed order calculations. 
We would like to point out that there is no complete NNLO calculation (i.e., third order in $a_s$) 
for the longitudinal coefficient functions available at this time. 
Therefore, only the first two non-vanishing logarithmic contributions can be resummed for the time being.
For this reason, the third entry for $\mathbb{C}_{L,g}^{S}$ in Tab.~\ref{tab:Ndependence_coefs} has to be 
deduced using analytic continuation ($\mathcal{AC}$) relations 
between DIS and SIA; see Refs.~\cite{ref:nnlo-kernel, Blumlein:2000wh} for details.
 
%%%%%%%%%%%%%%%%%%%%%%%%%%%%%%%%%%%%%
\subsection{Small-$z$ resummations \label{subsec:resum}}
%%%%%%%%%%%%%%%%%%%%%%%%%%%%%%%%%%%%%
%
The resummation of the first three towers of small-$z$ logarithms, summarized 
in Tables~\ref{tab:Ndependence_coefs} and \ref{tab:Ndependence_splittings}, 
was performed recently in Refs.~\cite{Vogt:2011jv,Kom:2012hd} in a formalism
based on all-order mass factorization relations and the general structure of
unfactorized structure functions in SIA. Explicit analytical results can be found for the choice 
$\mu=Q$. The corresponding LL and NLL expressios are known for quite some time \cite{ref:mueller,Bassetto:1982ma}
and have been derived by other means.
We have adopted the same framework based on mass factorization as in \cite{Vogt:2011jv,Kom:2012hd} 
and re-derived all results from scratch up to NNLL accuracy. We are in perfect agreement
with all of their expressions except for some obvious, minor typographical errors
\footnote{We noticed the following typographical errors in Ref.~\cite{Vogt:2011jv} which
should be corrected as follows:\\
Eq.~(2.12): $\left(\frac{67}{9} C_A - 4\zeta_2\right) \to \left(\frac{67}{9} - 4\zeta_2\right)$\\[0.5mm]
Eq.~(3.18) 1$^{\text{st}}$ line, last term: \mbox{$-\frac{38}{9}C_A^2 C_F n_f \to -\frac{38}{9}C_A C_F^2 n_f$}\\[0.5mm]
Eq.~(4.8) 2$^{\text{st}}$ line, last term: $-\frac{47}{9}C_F n_f^2 \to -\frac{47}{9}C_F^2 n_f $\\[0.5mm]
Eq.~(5.5) denominator: $9 (N-1)^{2n-2} \to 9 (N-1)^{2n-3}$
}. 
In this section, we will concisely summarize the main aspects of the calculation as we will extend
the obtained results to a general choice of scale $\mu\neq Q$ in the next subsection. 
 
One starts from the unfactorized structure functions using dimensional regularization. 
In our case, we choose to work in $d=4-2\eps$ dimensions. 
The unfactorized partonic structure functions can be written as
\be\label{eq:unfactstructfuncs}
\hat{\mathcal{F}}_{k,l}(N,a_s,\eps) = \sum_{i=q,g} C_{k,i}(N,a_s,\eps) \Gamma_{il}(N,a_s,\eps)\,,
\ee
with $k=L,T$ and $l=q,g$. 
We have introduced the $d$-dimensional coefficient functions $C_{k,l}$, 
which contain only positive powers in $\eps$,
\be\label{eq:ddimcoef}
C_{k,l}(N,a_s,\eps) = \delta_{kT}\delta_{l,q} + \sum_{i=1}^\infty a_s^i \sum_{j=0}^\infty \eps^j c_{k,l}^{(i,j)}(N)\,,
\ee
whereas the transition functions $\Gamma_{ij}$ include all IR/mass singularities, 
which are manifest in $1/\eps$ poles, i.e., they contain all negative powers of $\eps$. 
The transition functions are calculable order by order in $a_s$ by solving the equation
\be\label{eq:deftransfuncs}
\beta_d(a_s) \frac{\partial \Gamma_{ik}}{\partial a_s} \Gamma_{kj}^{-1} = P_{ij}^T\,.
\ee
Here, $\beta_d(a_s) = -\eps\, a_s - a_s^2 \sum_{i=0}^\infty \beta_i a_s^i$ denotes the
d-dimensional beta function of QCD. Eq.~\eqref{eq:deftransfuncs} can be derived from the time-like evolution 
equations and its solution reads
\ba\label{eq:gammaexplicit}
\bs\Gamma &=& \mathbb{1} - a_s \frac{\bs P^{T,(0)}}{\eps}\nn\\
&+& a_s^2\bigg[\frac{1}{2\eps^2}(\bs P^{T,(0)}+\beta_0)\bs P^{T,(0)}-\frac{1}{2\eps}\bs P^{T,(1)}\bigg]\nn\\
&+& a_s^3\bigg[-\frac{1}{6\eps^3} (\bs P^{T,(0)} + \beta_0)(\bs P^{T,(0)} + 2 \beta_0)\bs P^{T,(0)} +\nn\\
&& \frac{1}{6\eps^2}\bigg\{(\bs P^{T,(0)} + 2 \beta_0)\bs P^{T,(1)} + (\bs P^{T,(1)} + \beta_1)2\bs P^{T,(0)} \bigg\} \nn\\
&&- \frac{1}{3\eps} \bs P^{T,(2)} \bigg] + \mathcal{O}(a_s^4)
\ea
where 
\be\label{eq:PTmatrix}
\bs P^{T} \equiv \sum_{i=0}^\infty a_s^{i+1} \bs P^{T,(i)} \equiv \sum_{i=0}^\infty a_s^{i+1}  \Bigg (\begin{matrix}P^{T,(i)}_{qq} & P^{T,(i)}_{gq}\\[2mm] P^{T,(i)}_{qg} & P^{T,(i)}_{gg} \end{matrix} \Bigg )
\ee
is the $2\times 2$ matrix that contains the time-like singlet splitting functions.
Throughout this work, we use bold face characters to denote  $2\times 2$ matrices.
Since we are interested only in the small-$z$ regime, we take the small-$\nbar$ 
limit of the known coefficient and splitting functions in Eq.~\eqref{eq:unfactstructfuncs}.

Alternatively, one can express the unfactorized partonic structure functions in Eq.~(\ref{eq:unfactstructfuncs})
as a series in $a_s$, 
\be\label{eq:unfactseries}
\hat{\mathcal{F}}_{k,l}(N,a_s,\eps) = \sum_{n}a_s^n \hat{\mathcal{F}}_{k,l}^{(n)}(N,a_s,\eps)\,.
\ee
The key ingredient to achieve the resummations of the leading small-$\nbar$ contributions, 
which is the main result of \cite{Vogt:2011jv}, is 
the observation that the ${\cal{O}}(a_s^n)$ contribution in Eq.~\eqref{eq:unfactseries} may be written as
\ba\label{eq:vogtmasterformula}
\hat{\mathcal{F}}_{k,l}^{(n)}(N,a_s,\eps) =
\eps^{\delta_{kL}+\delta_{lq}+1-2n}\sum_{i=0}^{n-1-\delta_{lq}} \frac{1}{\nbar-2(n-i)\eps}\nn\\
\times\left(A_{k,l}^{(i,n)} + \eps B_{k,l}^{(i,n)} + \eps^2  C_{k,l}^{(i,n)} + \ldots \right)\,.\nn\\~
\ea  
Each of the coefficients $A$, $B$, and $C$ is associated with a different 
logarithmic accuracy of the resummation, i.e., LL, NLL, and NNLL, respectively.

By equating Eqs.~\eqref{eq:unfactstructfuncs} and \eqref{eq:unfactseries},
one obtains a system of equations which may be solved recursively order by order in $a_s$. 
The small-$z$ (small-$\nbar$) limits of the fixed order results are needed here 
as initial conditions for the first recursion. Since these results are only known up to NNLO accuracy,
resummations are limited for the time being to the first three towers 
listed in Tables~\ref{tab:Ndependence_coefs} and \ref{tab:Ndependence_splittings}. 
%They are listed in \cite{Vogt:2011jv} up to a typo listed in Appendix[TO DO: reference to Daniele Appendix].
At each order $n$, this procedure then yields expressions
for $c_{k,l}^{(n,m)},\; P_{ij}^{T,\,(n-1)},\; A_{k,l}^{(m,n)},\; B_{k,l}^{(m,n)}$, and $C_{k,l}^{(m,n)}$. 

Note that up to NNLL accuracy only $\beta_0$ is needed in Eq.~\eqref{eq:gammaexplicit}. 
All terms proportional $\beta_{i\ge 1}$ will generate subleading contributions and, hence, can be discarded.
For instance, when initiating the recursive solution, $\bs P^{T,(0)}$ and $\bs P^{T,(1)}$ are known from 
fixed order calculations, and $\bs P^{T,(2)}$, that appears at $\mathcal{O}(a_s^3)$ in Eq.~\eqref{eq:gammaexplicit},
is the unknown function that is being determined.
The NNLL contribution for, say, $P_{gg}^{T,\,(2)}$ is $\propto 1/ \bar N ^2$, cf.\ Table~\ref{tab:Ndependence_splittings},
whereas the highest inverse power of $\bar N$ in the term $\beta_1 P_{gg}^{T,\,(0)}$ 
appearing in the curly brackets of Eq.~\eqref{eq:gammaexplicit} 
is $\propto 1/\bar N$ and, thus, beyond NNLL accuracy.

After solving the system of equations algebraically using \textsc{Mathematica} \cite{ref:mathematica}, 
we find expressions for $c_{k,l}^{(n,0)}$, and $P_{ij}^{T,(n)}$. Since the 
coefficient functions and the splitting functions both have a perturbative expansion in $a_s$,
\be
P_{ij} ^{T}= \sum_{n=0}^{\infty} a_s^{n+1} P_{ij}^{T,\,(n)}
\ee
and
\be
\mathbb{C}_{k,l}^{\text{S}} = \sum_{n=0}^{\infty} a_s^{n} c_{k,l}^{(n,0)}
\ee
one can eventually deduce a closed expression for resummed splitting functions and coefficient functions as listed in \cite{Kom:2012hd}. As mentioned above, we fully agree with these results up to the typographical errors listed
in the footnote.

%%%%%%%%%%%%%%%%%%%%%%%%%%%%%%%%%%%%%
\subsection{Resummed scale dependence \label{subsec:scales}}
%%%%%%%%%%%%%%%%%%%%%%%%%%%%%%%%%%%%%
%
All calculations presented so far, including Refs.~\cite{Vogt:2011jv,Kom:2012hd},
have been performed by identifying, for simplicity, the renormalization and factorization scales 
with the hard scale $Q$, i.e., by setting $\mu_F = \mu_R = \mu = Q$.
However, it is well known that the resummation procedure should 
not only yield more stable results but should also lead to a better control 
of the residual dependence on the unphysical scales $\mu_F$ and $\mu_R$
that arises solely from the truncation of the perturbative series.
Hence, for our subsequent studies of the phenomenological impact of the small-$z$ resummations
on the extraction of FFs from SIA data it is imperative to reintroduce the dependence on the scales
$\mu_F$ and $\mu_R$ in the resummed expressions. This is the goal of this section.
In what follows, we reinstate the scale dependence with two different, independent methods. 
We find full agreement between the two approaches. 

Firstly, we consider a 
renormalization group approach; see also Ref.~\cite{vanNeerven:2000uj}. The dependence of the coefficient functions on the factorization scale $\mu_F$ can be expressed as
\be\label{eq:coefscales}
\mathbb{C}_{k,l}^{\text{S}}(N,a_s,L_M) = \sum_{i=0}^\infty a_s^i  \left(  c_{k,l}^{(i)}(N) + \sum_{m=1}^{i}
\tilde{c}_{k,l}^{(i,m)}(N) L_M^m \right)\,,
\ee
with $L_M \equiv \log\frac{Q^2}{\mu_F^2}$. The coefficients $c_{k,l}^{(i)}\equiv \tilde{c}_{k,l}^{(i,0)}$ 
are the finite (i.e., $\eps$ independent) coefficients as given in Eq.~\eqref{eq:ddimcoef}. 
The $\tilde{c}_{k,l}^{(i,m)}$ can be calculated order by order in $a_s$ by solving a set of 
renormalization group equations (RGEs). 
These equations can be obtained by requiring that $\frac{d}{d\log\mu_F^2} F_k\overset{!}{=} 0$, 
where \mbox{$F_k \equiv \sum_l \mathbb{C}_{k,l} D_l$} (see Eq.~\eqref{eq:FTFL} for the definition
of these structure functions in $z$ space), 
which leads to
\begin{widetext}
\be\label{eq:rge1}
\bigg[\bigg\{\frac{\partial}{\partial \log\mu_F^2}  +  
\beta(a_s)\frac{\partial}{\partial a_s} \bigg\}\delta_{lm} + 
P^{T}_{lm}(N) \bigg]\mathbb{C}_{k,m}^{\text{S}}(N,a_s,L_M) = 0\,.
\ee
Here, the sum over $m=q,g$ is left implicit. For the sake of better readability, 
we drop the arguments of all functions for now. From~(\ref{eq:rge1}), the following recursive formula can be obtained
\be
\tilde{c}_{k,l}^{(i,m)} = \frac{1}{m} \sum_{w=m-1}^{i-1} \tilde{c}_{k,j}^{(w,m-1)} 
\left(P_{lj}^{T,\, (i-w-1)} - w \beta_{i-w-1} \delta_{jl} \right)\,.
\ee
\end{widetext}
Again, the sum over $j=q,g$ is implicitly understood.
Up to NNLO accuracy, we obtain the same results as given in \cite{ref:nnlo-sia}.

If one now plugs in the small-$\nbar$ results for the splitting and coefficient functions, one can compute 
the coefficients $\tilde{c}_{k,l}^{(n,m)}$ up to any order $n$ and identify the leading three towers
of $1/\nbar$ in Eq.~\eqref{eq:coefscales}, i.e., the LL, NLL, and NNLL contributions. 
At order $n$ we find at LL accuracy
\be
\label{eq:CLLscale}
\mathbb{C}_{k,g}^{\text{S},\LL,(n)} = c_{k,g}^{\LL,(n)}\,.
\ee
Thus, no improvement of the scale dependence is achieved by a LL resummation 
(recall that resummation in the quark sector only starts at NLL accuracy). 
The full $L_M$ dependence is given by the fixed-order expressions, which have to be matched to the 
resummed result for all practical purposes. 
As usual, the matching of a resummed observable $T^{\text{res}}$ to its $\text{N}^\kappa\text{LO}$ 
fixed-order expression $T^{\text{N}^\kappa\text{LO}}$ is performed according to
the prescription schematically given by
\be\label{eq:matching}
 T^{\text{matched}}= T^{\text{N}^\kappa\text{LO}}+T^{\text{res}}-\left. T^{\text{res}}\right|_{\mathcal{O}(a_s^\kappa)}\,.
\ee
Here, $\left.T^{\text{res}}\right|_{\mathcal{O}(a_s^\kappa)}$ denotes the expansion in $a_s$ of $T^{\text{res}}$ up to order $\mathcal{O}(a_s^\kappa)$. 

\begin{widetext}
Likewise, at NLL accuracy one obtains the following results 
\ba
\label{eq:CNLLscale}
\mathbb{C}_{T,g}^{\text{S},\NLL,(n)} = c_{T,g}^{\NLL,(n)} + 
L_M \Bigg\{ P_{gq}^{T\;\LL, (n-1)}  
+ \sum_{j=0}^{n-2} c_{T,g}^{\LL, (n-1-j)} P_{gg}^{T\;\LL, (j)}   \Bigg\} \,,
\ea
\ba
\label{eq:CNLLscale2}
\mathbb{C}_{L,g}^{\text{S},\NLL,(n)} = c_{L,g}^{\NLL,(n)} &+& 
L_M \sum_{j=0}^{n-2} c_{L,g}^{\LL, (n-1-j)} P_{gg}^{T\;\LL, (j)}\nn\\  
\ea
and
\ba
\mathbb{C}_{T,q}^{\text{S},\NLL,(n)} &=& c_{T,q}^{\NLL,(n)}\,,\\
\mathbb{C}_{L,q}^{\text{S},\NLL,(n)} &=& c_{L,q}^{\NLL,(n)}\, .
\ea
The scale dependent terms $\propto L_M$ enter here for the first time in the gluonic sector, Eqs~(\ref{eq:CNLLscale})
and (\ref{eq:CNLLscale2}), and are expressed in terms of LL quantities. 
Due to the fact that the quark coefficient functions are subleading, they still do not carry any 
scale dependence at NLL. 
Finally, at NNLL accuracy one finds
\ba
\mathbb{C}_{T,g}^{\text{S},\NNLL,(n)} = c_{T,g}^{\NNLL,(n)} &+& L_M \Bigg\{P_{gq}^{T\;\NLL, (n-1)} - (n-1)\beta_0 c_{T,g}^{\LL, (n-1)} + \sum_{j=0}^{n-3}c_{T,q}^{\NLL, (n-1-j)} P_{gq}^{T\;\LL, (j)} \nn\\
 &&~~ + \sum_{j=0}^{n-2}\Bigg( c_{T,g}^{\LL, (n-1-j)} P_{gg}^{T\;\NLL, (j)}+c_{T,g}^{\NLL, (n-1-j)} P_{gg}^{T\;\LL, (j)}\Bigg)\Bigg\} \nn\\
&+& \frac{L_M^2}{2}\Bigg[\sum_{j=0}^{n-2} P_{gq}^{T\;\LL, (n-2-j)} P_{gg}^{T\;\LL, (j)}  + \sum_{i=0}^{n-3} ~\sum_{j=0}^{n-2-i} c_{T,g}^{\LL, (n-2-i-j)} P_{gg}^{T\;\LL, (i)}P_{gg}^{T\;\LL, (j)}\Bigg] \,,
\ea
\ba
\mathbb{C}_{L,g}^{\text{S},\NNLL,(n)} =  c_{L,g}^{\NNLL,(n)} &+& L_M \Bigg\{-(n-1)\beta_0 c_{L,g}^{\LL, (n-1)}  +\sum_{j=0}^{n-2} \Bigg( c_{L,g}^{\LL, (n-1-j)} P_{gg}^{T\;\NLL, (j)}+c_{L,g}^{\NLL, (n-1-j)} P_{gg}^{T\;\LL, (j)}\Bigg) \nn\\
&&~~ +\sum_{j=0}^{n-2} c_{L,q}^{\NLL, (n-1-j)} P_{gq}^{T\;\LL, (j)} \Bigg\} +  \frac{L_M^2}{2} \sum_{i=0}^{n-3} ~\sum_{j=0}^{n-3-i} 
c_{L,g}^{\LL, (n-2-i-j)} P_{gg}^{T\;\LL, (i)}P_{gg}^{T\;\LL, (j)}\;,
\ea
\ba
\mathbb{C}_{T,q}^{\text{S},\NNLL,(n)} &=& c_{T,q}^{\NNLL,(n)} + L_M \Bigg\{P_{qq}^{T\;\NLL, (n-1)} (1-\delta_{n,2}) + \sum_{j=0,j\neq1}^{n-1} c_{T,g}^{\LL, (n-1-j)} P_{qg}^{T\;\NLL, (j)} \Bigg\} \;,
\ea
and
\ba
\label{eq:CNNLLscale}
\mathbb{C}_{L,q}^{\text{S},\NNLL,(n)} &=& c_{L,q}^{\NNLL,(n)} + 
L_M \sum_{j=0,j\neq1}^{n-2} c_{L,g}^{\LL, (n-1-j)} P_{qg}^{T\;\NLL, (j)}\nn\\
\ea
\end{widetext}

It should be noticed that by the subscripts LL, NLL, and NNLL in Eqs.~\eqref{eq:CLLscale} 
and~\eqref{eq:CNLLscale}-\eqref{eq:CNNLLscale}, we denote {\em only} those
contributions in $1/\bar N$ specific to the tower at LL, NLL, or NNLL accuracy, respectively. 
This means, for instance, that the full next-to-next-to-leading logarithmic expression 
at some given order $n$ in the $a_s$ perturbative expansion of 
$\mathbb{C}_{k,l}^{\text{S}}$ in Eq.\eqref{eq:coefscales} 
will be always given by the {\em sum} of the individual LL, NLL, and NNLL contributions.
As one may expect from the fixed-order results, the scale dependence at N$^m$LL is expressed entirely
in terms of the resummed expressions at N$^k$LL with $k<m$.
Since the resummed results are known up to NNLL accuracy, we may, in principle, extend our calculations 
to fully predict the scale dependent terms at N$^3$LL. These findings are consistent with the 
scale dependence of fixed-order cross sections.
Finally, for all practical purposes, as we shall see below, it is numerically adequate to have explicit results
for each tower up to sufficiently high order in $n$, say, $n=20$, in lieu of a closed 
analytical expression for the resummed series as was provided for the case $\mu=Q$ in 
Refs.~\cite{Vogt:2011jv,Kom:2012hd}.
 
We may now reintroduce the renormalization scale dependence as well by following the straightforward 
steps outlined in Ref.~\cite{ref:nnlo-sia}. 
In practice, this amounts to replacing all couplings $a_s$ in the expressions given above according to
\be
a_s(\mu_F^2)=a_s(\mu_R^2)\left(1+a_s(\mu_R^2)\beta_0\log{\frac{\mu_R^2}{\mu_F^2}}+\mathcal{O}(a_s^2)\right)\,.
\ee
In a second step one needs to re-expand all results in terms of $a_s(\mu_R^2)$ which leads 
to additional logarithms of the type $L_R\equiv \log(\mu_R^2/\mu_F^2)$. 
In our phenomenological studies below we will study, however, only the case $\mu_F=\mu_R\neq Q$ and, hence,
we do not pursue the $L_R$ dependence any further.

The second approach we adopt to recover the scale dependence of the SIA coefficient functions
obtained in Sec.~\ref{subsec:resum}
is based on the all-order mass factorization procedure. After removing the ultraviolet (UV)
singularities from the bare partonic structure functions $\hat{\mathcal{F}}_{k,l}$ (which have
been computed directly from Feynman diagrams) by a suitable renormalization procedure,
the remaining final-state collinear/mass singularities have to be removed by mass factorization 
\be\label{eq:massfact}
\tilde{\mathcal{F}}_{k,l} = \mathbb{C}_{k,i} \otimes \tilde{\Gamma}_{li}\,.
\ee
Here, all singularities are absorbed into the transition functions $\tilde{\Gamma}_{li}$ while the coefficient functions
$\mathbb{C}_{k,i}$ are finite. We have labeled the quantities in Eq.~\eqref{eq:massfact}
with a tilde to show that they contain the full dependence on all scales.

We may thus proceed in the following way: first, we ``dress'' the transition functions and partonic structure functions 
in Eq.~(\ref{eq:unfactstructfuncs}) with the appropriate scale dependence, i.e., we substitute 
$a_s \to a_s\cdot (\mu_F^2/\mu^2)^{-\eps}$ in the $\Gamma_{ij}$ and 
$a_s \to a_s\cdot (Q^2/\mu^2)^{-\eps}$ in the $\hat{\mathcal{F}}_{k,l}$,
where the mass parameter $\mu$ stems from adopting dimensional regularization.
As a next step, we go back to the unrenormalized
expressions, where we assume that the renormalization was performed at the scale 
$\mu_F^2$ and $Q^2$, respectively. 
Afterwards, we perform renormalization again, but now at a different scale $\mu_R^2$. 
Schematically, this amounts to
\be\label{eq:getgammatilde}
\tilde{\Gamma}_{ij} = R_{\mu^2}^{\mu_R^2}\left[(R_{\mu^2}^{\mu_F^2})^{-1}
\left[\Gamma_{ij}(a_s \to a_s\cdot (\mu_F^2/\mu^2)^{-\eps}) \right] \right]
\ee
and
\be\label{eq:getftilde}
\tilde{\mathcal{F}}_{k,l} = R_{\mu^2}^{\mu_R^2}\left[(R_{\mu^2}^{Q^2})^{-1}\left[\mathcal{F}_{k,l}(a_s \to a_s\cdot (Q^2/\mu^2)^{-\eps}) \right] \right]\,.
\ee
Here, we are using the following notation: with $R_{\mu^2}^{\mu_R^2}[\hat{f}(\hat{a}_s)] = f[a_s(\mu_R^2)]$ 
we denote the renormalization of a bare quantity $\hat{f}(\hat{a}_s)$ which, as indicated, depends on the
unrenormalized, bare coupling $\hat{a}_s$. This procedure yields a renormalized quantity $f[a_s(\mu_R^2)]$, 
which now depends on the physical coupling $a_s(\mu_R^2)$. 
The renormalization procedure $R_{\mu^2}^{\mu_R^2}$ is performed by
replacing the bare coupling with
\be
\hat{a}_s = a_s(\mu_R^2) Z(\mu_R^2,\mu^2)
\ee
where we have introduced the renormalization constant
\be
\label{eq:renconst}
Z(\mu_R^2,\mu^2) \equiv \left[1 - a_s(\mu_R^2)\cdot\left(\frac{\mu_R^2}{\mu^2}\right)^{-\eps} \frac{\beta_0}{\eps} + \mathcal{O}(a_s^2)  \right]\,.
\ee
Analogously, $(R_{\mu^2}^{\mu_R^2})^{-1}[f[a_s(\mu_R^2)]] = \hat{f}(\hat{a}_s)$ 
performs the inverse operation, i.e., it translates the renormalized quantity $f(a_s(\mu_R^2))$
back to the corresponding bare quantity $\hat{f}(\hat{a}_s)$.
This is achieved by replacing the renormalized coupling with   
\be
a_s(\mu_R^2) = \hat{a}_s \hat{Z}(\mu_R^2,\mu^2)\;,
\ee
where the ``inverse" renormalization constant reads
\be
\hat{Z}(\mu_R^2, \mu^2) \equiv \left[1 + \hat{a}_s\cdot\left(\frac{\mu_R^2}{\mu^2}\right)^{-\eps} \frac{\beta_0}{\eps} + \mathcal{O}(\hat{a}_s^2)  \right]\,.
\ee
The latter can be obtained from Eq.~\eqref{eq:renconst} by a series reversion. 
After substituting Eqs.~\eqref{eq:getgammatilde} and \eqref{eq:getftilde} into Eq.~\eqref{eq:massfact} one can  
solve the latter equation for the coefficients $\mathbb{C}_{k,i}$, 
which now exhibit the full dependence on $\mu_R$ and $\mu_F$.

In order to generate the renormalization constant $Z$ in Eq.~(\ref{eq:renconst})
at each order $n$ in an expansion in $a_s$ with the maximal precision available at this time (i.e., 
up to terms proportional to $\beta_i$, $i\le2$), 
we adopt renormalization group techniques. The general form of the renormalization constant reads
\be
Z = 1 +\sum_{k=1}^\infty a_s^k \sum_{l=1}^k \frac{f_{k,l}}{\eps^l}
\ee
and may also be written as
\be
Z = 1 + \sum_{l=1}^\infty \frac{g_l(a_s)}{\eps^l}
\ee
where $g_l(a_s) = \sum_{k=l}^\infty a_s^k f_{k,l}$ is a power series in $a_s$ with $l$ being the lowest power. 
Using the RGE it is possible to derive a recursive formula for this power series,
\be\label{eq:glrec}
g_{k+1}^\prime(a_s) = g_1^\prime(a_s) \frac{d (a_s g_k(a_s))}{da_s}\,.
\ee
Here the prime denotes a derivative with respect to $a_s$. 
Hence, we obtain $g_{k+1}(a_s)$ by integration of Eq.~\eqref{eq:glrec}. 
From the expression of the renormalization constant up to $a_s^3$, see, for example Ref.~\cite{Moch:2005id}, 
we obtain as initial conditions
\be
f_{1,1} = -\beta_0,~~~~~f_{2,1} = -\frac{\beta_1}{2},~~~~~f_{3,1} = -\frac{\beta_2}{3}\,.
\ee
As already stated above, only terms proportional to $\beta_0$ are relevant up to NNLL accuracy.

%%%%%%%%%%%%%%%%%%%%%%%%%%%%%%%%%%%%%
\subsection{Solution to the time-like evolution equation with a resummed kernel\label{subsec:evol}}
%%%%%%%%%%%%%%%%%%%%%%%%%%%%%%%%%%%%%
% 
The dependence of the gluon and $N_f$ quark and antiquark FFs on the 
factorization scale $\mu_F$ is governed by a set of $2N_f+1$ RGEs,
which are the time-like counterparts of the well-known equations
pertinent to the scale evolution of PDFs \cite{ref:DGLAP}.
Schematically, they can be written as
\begin{equation}
\label{eq:evolution}
\frac{\partial}{\partial\ln\mu^2}D^h_i(z,\mu^2)= \sum_j P^T_{ji}(z,\mu^2)\otimes D^h_j\left(z,\mu^2\right)\, ,
\end{equation}
with $i,j=q,\bar{q},g$. For simplicity, we have set $\mu_R=\mu_F=\mu$ as in Sec.~\ref{subsec:xsec}.
The $i\to j$ splitting functions $P^T_{ji}(z,\mu^2)$ obey a perturbative expansion in $a_s$,
\begin{equation}
\label{eq:splittingexp}
P^{T}_{ji} = a_s P_{ji}^{T,(0)} + a_s^2 P_{ji}^{T,(1)} +  a_s^3 P_{ji}^{T,(2)} + \ldots \, ,
\end{equation}
where we have suppressed the arguments $z$ and $\mu^2$. As discussed extensively in \cite{Anderle:2015lqa},
up to a minor ambiguity concerning the off-diagonal splitting kernel $P_{qg}^{T,(2)}$, the
expansion (\ref{eq:splittingexp}) is known up to NNLO accuracy \cite{ref:nnlo-kernel}, i.e., ${\cal O}(a_s^3)$. 
Presumably, this remaining uncertainty, which stems from adopting {$\mathcal{AC}$} relations
on the known NNLO space-like results, is numerically irrelevant for all phenomenological applications;
see Ref.~\cite{ref:Pijdirectcalculation} for the status of an ongoing direct calculation
of the three-loop time-like kernels.

Instead of the fixed-order expressions defined in Eq.~(\ref{eq:splittingexp}),
we shall consider the resummed results for the splitting functions $P_{jl}^{T\;\text {N}^\kappa\text{LL}}$ 
as discussed in Sec.~\ref{subsec:resum} and listed in Ref.~\cite{Vogt:2011jv, Kom:2012hd}.
The obey a similar expansion in $a_s$ as in  Eq.\eqref{eq:splittingexp}, which reads
\begin{equation}
\label{eq:splittingresumexp}
P_{ji}^{T\; \text {N}^\kappa\text{LL}}  = 
\sum^\infty_{n=0} a_s^{n+1} P_{ji}^{T\; \text {N}^\kappa\text{LL},(n)}  \;,
\end{equation}
where each term $P_{ji}^{T\; \text {N}^\kappa\text{LL},(n)}$ in (\ref{eq:splittingresumexp}) is, 
in principle, known up to NNLL accuracy, i.e., for $\kappa=0$, 1, and 2.
 
Before extending the technical framework to solve Eq.~\eqref{eq:evolution} in Mellin moment space
to the resummed case, we briefly summarize hereinafter the methods and strategies
used in the fixed-order approach as they remain relevant.
Here, we closely follow Ref.~\cite{ref:pegasus}
and the notation adopted in a recent analysis of pion FFs at NNLO accuracy \cite{Anderle:2015lqa}.

For the singlet sector, Eq.~\eqref{eq:evolution} translates into 
two coupled integro-differential equations, which read
\begin{equation}
\label{eq:singletevol}
\frac{d}{d\ln\mu^2} \Bigg (\begin{matrix}D^h_{\Sigma} \\[2mm] D^h_g
\end{matrix} \Bigg ) =  \Bigg (\begin{matrix}P^T_{qq} & 2N_f P^T_{gq}\\[2mm] \frac{1}{2N_f}P^T_{qg} & P^T_{gg} \end{matrix} \Bigg ) \otimes \Bigg (\begin{matrix}D^h_{\Sigma} \\[2mm] D^h_g
\end{matrix} \Bigg ) \, ,
\end{equation}
where 
\be \label{eq:singlet}
D^h_{\Sigma} \equiv \sum_{q}^{N_f}(D^h_{q}+D^h_{\bar q}) 
\ee
is the singlet flavor combination, i.e., $N_f$ times the combination $D^h_S$, defined 
in (\ref{eq:sing}), that appears in the SIA cross section (\ref{eq:nnlostructure}), 
and $D^h_g$ denotes the gluon FF.

The remaining $2N_f-1$ equations can be fully decoupled
by choosing the following, convenient non-singlet combinations of FFs: 
\ba \label{eq:NSPM}
D^{h,\pm}_{\mathrm{NS},l} &\equiv& \sum_{i=1}^k (D^h_{q_i}\pm D^h_{\bar q_i}) - k (D^h_{q_k}\pm D^h_{\bar q_k})\, , \\ 
\label{eq:NSVAL}
D^h_{\mathrm{NS},v} &\equiv& \sum_{q}^{N_f}(D^h_{q}-D^h_{\bar q})\,.
\ea
In Eq.~(\ref{eq:NSPM}), we have $l=k^2-1$, $k=2,\ldots,N_f$, and the subscripts $i,k$ 
were introduced to distinguish different quark flavors.
Each combination in Eqs.~\eqref{eq:NSPM} and \eqref{eq:NSVAL} evolves independently 
with the following NS splitting functions \cite{ref:nnlo-kernel}
\ba
\label{eq:splittingNSPM}
P^{T,\pm}_{\mathrm{NS}}&=&P^{T,v}_{qq}\pm P^{T,v}_{q\bar q}\,,\\
P^{T,v}_{\mathrm{NS}}&=& 
%P^{T,v}_{qq}-  P^{T,v}_{q\bar q}+N_f(P^{T,s}_{qq}-P^{T,s}_{q\bar q})\\
\label{eq:splittingNSVAL}
%&\equiv&
P^{T,-}_{\mathrm{NS}}+P^{T,s}_{\mathrm{NS}} \, ,
\ea
respectively, and one has the following relation for $P_{qq}^T$ that enters in Eq.~(\ref{eq:singletevol})
\begin{eqnarray}
\label{eq:splittingS}
P^{T}_{qq} 
= P^{T,+}_{\mathrm{NS}}+P^{T,ps} \, .
\end{eqnarray}
Similar to the space-like case, one finds $P^{T,v}_{q\bar q} = P^{T,s}_{\mathrm{NS}} = P^{T,ps} = 0$ and
$P^{T,s}_{\mathrm{NS}}=0$ at LO and NLO, respectively. Hence, three
NS quark combinations that evolve differently first appear at NNLO accuracy \cite{ref:nnlo-kernel}. 
After the evolution is performed, i.e., the singlet and the $(2N_f-1)$ non-singlet equations are solved,
the individual $D_q^h$ and $D_{\bar{q}}^h$ can
be recovered from Eqs.~(\ref{eq:singlet}), (\ref{eq:NSPM}), and (\ref{eq:NSVAL}).
Likewise, any combination relevant for a cross section calculation can be computed,
such as those used in the factorized expression for SIA given in Eq.~(\ref{eq:nnlostructure}).

As for the resummations of the small-$z$ logarithms in Secs.~\ref{subsec:resum} and \ref{subsec:scales},
it is most convenient to solve the set of evolution equations in Mellin $N$ space,  
exploiting the fact that all convolutions $\otimes$ turn into simple products in moment space.
Hence, one can rewrite all evolution equations as ordinary differential equations. 
Schematically, one finds
\ba
\label{eq:mtevolution}
\frac{\partial \boldsymbol D^h(N,a_s)}{\partial a_s}&=&-\frac{1}{a_s}\bigg[\boldsymbol R_0(N)+\sum^\infty_{k=1}a_s^k\boldsymbol R_k(N)\bigg]\boldsymbol D^h(N,a_s)\, , \nn \\
\ea
where the characters in boldface indicate that we are dealing in general with 
$2\times 2$ matrix-valued equations, cf.\ Eq.~(\ref{eq:singletevol}). 
For the NS combinations (\ref{eq:NSPM}) and (\ref{eq:NSVAL}), 
Eq.~(\ref{eq:mtevolution}) reduces to a set of independent partial differential equations,
which are straightforward to solve, and we do not discuss them here.

The $\boldsymbol R_k$ in~(\ref{eq:mtevolution}) are defined recursively by
\be \label{eq:Rmatrix}
\boldsymbol R_0\equiv \frac{1}{\beta_0} \widetilde{\boldsymbol P}^{T,(0)}\;,\;\; \boldsymbol R_k \equiv \frac{1}{\beta_0}\widetilde{\boldsymbol P}^{T,(k)}-\sum^k_{i=1}b_i\boldsymbol R_{k-i}\;,
\ee
where $\widetilde{\boldsymbol P}^{T,(k)}(N)$ is the $k$-th term in the perturbative expansion of the
$2\times 2$ matrix of the $N$-moments of the singlet splitting functions
\be
\label{eq:PTtilde}
\widetilde{\boldsymbol P}^{T}(N)=\Bigg (\begin{matrix}P^T_{qq}(N) & 2N_f P^T_{gq}(N)\\[2mm] \frac{1}{2N_f}P^T_{qg}(N) & P^T_{gg}(N) \end{matrix} \Bigg )\,.
\ee
Note that here and in Eq.~(\ref{eq:singletevol}),
the off-diagonal entries of the matrix $\widetilde{\boldsymbol P}^{T}$ differ from the ones 
of $\boldsymbol P^{T}$ in Eq.~\eqref{eq:PTmatrix} by factors $2 N_f$ and $1/2 N_f$. 
This is simply due to the different definitions used for the singlet combination 
in the evolution \eqref{eq:singletevol} and in the calculation of the SIA cross section~\eqref{eq:nnlostructure}, 
c.f.\ Eqs.~ \eqref{eq:sing} and \eqref{eq:singlet}.
In addition, we have introduced $b_i \equiv \beta_i/\beta_0$, where $\beta_k$ denote 
the expansion coefficients of the QCD $\beta$-function;
see Ref.~\cite{ref:betafct} for the explicit expressions up to NNLO, i.e., $\beta_2$.

Due to the matrix-valued nature of Eq.~(\ref{eq:mtevolution}), no unique closed
solution exists beyond the lowest order approximation. Instead, it can be written 
as an expansion around the LO solution, 
$(a_s/a_0)^{-\boldsymbol R_0(N)} \boldsymbol D^h(N,a_0)$. Here, $a_0$ is the value of $a_s$ at the
initial scale $\mu_0$, where the non-perturbative input $\boldsymbol D^h(N,a_0)$ is specified from a fit to data.
More explicitly, this expansion reads
\begin{eqnarray}
\label{eq:mtgeneralsolution}  
\boldsymbol D^h(N,a_s)&=&\bigg[1+\sum^\infty_{k=1}a_s^k\,\boldsymbol U_k(N)\bigg]
\bigg(\frac{a_s}{a_0}\bigg)^{-\boldsymbol R_0(N)} \nonumber\\[2mm]
&\times & \bigg[1+\sum^\infty_{k=1}a_s^k\,\boldsymbol U_k(N)\bigg]^{-1}\boldsymbol D^h(N,a_0)\;.\hspace*{5mm}
\end{eqnarray}
The evolution matrices $\boldsymbol U_k$ are again defined recursively by the commutation relations
\begin{eqnarray}
\label{eq:umatrix}
[\boldsymbol U_k,\boldsymbol R_0]=\boldsymbol R_k+\sum_{i=1}^{k-1}\boldsymbol R_{k-1}\boldsymbol U_i+k\boldsymbol U_k\;.
\end{eqnarray}
When examining Eq.~\eqref{eq:mtgeneralsolution} more closely, it turns out 
that a fixed-order solution at N$^m$LO accuracy
is not unambiguously defined. A certain degree of freedom still remains in choosing the details on how to
truncate the series at order $m$.  
For example, suppose the perturbatively calculable quantities 
$\widetilde{\boldsymbol P}^{T,(k)}$ and $\beta_k$ are available up to a certain order $k=m$.
One possibility is to expand Eq.~(\ref{eq:mtgeneralsolution}) in $a_s$ and strictly keep only 
terms up to $a_s^m$. This defines what is usually called the \emph{truncated solution} in Mellin moment space.

However, given the iterative nature of the $\boldsymbol R_k$ in Eq.~(\ref{eq:Rmatrix}), 
one may alternatively calculate the $\boldsymbol R_k$ and, 
hence, the $\boldsymbol U_k$ in Eq.~(\ref{eq:umatrix}) for any $k>m$ from the known results 
for $\widetilde{\boldsymbol P}^{T,(k)}$ and $\beta_k$ up to $k=m$. 
Any higher order $\widetilde{\boldsymbol P}^{T,(k>m)}$ and $\beta_{k>m}$ are simply set to zero.
Taking into account all the thus constructed $\boldsymbol U_k$
in Eq.~(\ref{eq:mtgeneralsolution}) defines the so-called \emph{iterated solution}. 
This solution is important as it mimics the results that are obtained when solving Eq.~(\ref{eq:evolution}) 
directly in $z$-space by some iterative, numerical methods. 
It should be stressed that both choices 
are equally valid as they only differ by terms that are of order $\mathcal O (a_s^{m+1})$.

The simplest way of extending the fixed-order framework outlined above 
to the resummed case is to take the \emph{iterated solution}. 
However, instead of setting contributions beyond the fixed order to zero, we use 
the resummed expressions.
One can define a $\text{N}^{m}\text{LO}$+$\text{N}^\kappa\text{LL}$ resummed ``matched solution'' by defining the
$k$-th term of the splitting matrix which appears in Eq.~\eqref{eq:Rmatrix} as follows:
\begin{equation}
\widetilde{\boldsymbol P}^{T,(k)} \equiv 
\begin{cases} 
\widetilde{\boldsymbol P}^{T\; \text{FO},(k)} & k \leq m \\
\widetilde{\boldsymbol P}^{T\; \text {N}^\kappa\text{LL},(k)} & k > m\,.
\end{cases} 
\end{equation}
In other words, the full fixed-order expressions $\widetilde{\boldsymbol P}^{T\; \text{FO},(k)}$ for $k \leq m$ are kept 
in $\boldsymbol R_k$, whereas we use the resummed expressions for $k > m$. 
This iterated and matched solution is the one implemented in our numerical code and will be used
in Sec.~\ref{sec:fit} for all our phenomenological studies. 
For the range of $z$-values covered by the actual data sets considered in this paper, 
only the terms up to $k=20$ are indeed numerically relevant as we shall discuss further 
in Sec.~\ref{subsec:num}. However, when evolving the FFs in scale 
with such an extended iterative solution, one finds that momentum conservation is broken 
to some extent due to missing sub-leading terms in the evolution kernels. 

In fact, total momentum conservation for FFs is expressed by the sum rules
for combinations of splitting functions, see, e.g.\ Ref.~\cite{Altarelli:1977zs}. 
\ba
\int_0^1 dx\, x \left(P_{qq}^T(x) + P_{gq}^T(x)\right) &=& 0\,,\nn\\
\int_0^1 dx\, x \left(P_{gg}^T(x) + P_{qg}^T(x)\right) &=& 0\,.
\ea
In terms of Mellin moments, these relations read
\ba
P_{qq}^T(N=2) + P_{gq}^T(N=2) &=& 0\,,\label{eq:mellin_sumrule1}\\
P_{gg}^T(N=2) + P_{qg}^T(N=2) &=& 0\,.\label{eq:mellin_sumrule2}
\ea
These sum rules are satisfied, i.e., built into the kernels, at any given fixed order.

In the case of the iterated and matched solution we use in our numerical implementation, 
the sum rules in Eqs.~\eqref{eq:mellin_sumrule1} and \eqref{eq:mellin_sumrule2} deviate 
from zero only about a few $\permil$ which is perfectly tolerable. 
We note, that in calculations of the SIA cross section, we also adopt the matching procedure 
for the relevant resummed coefficient functions as specified in Eq.~(\ref{eq:matching}). 

However, when evaluating the sum rules without matching, the sums in \eqref{eq:mellin_sumrule1} and 
\eqref{eq:mellin_sumrule2} yield the approximate values $0.05$ and $0.1$, respectively, which is, of course,
not acceptable.

We would like to point out that a NLO \emph{truncated} + resummed solution has been proposed in 
Ref.~\cite{Blumlein:1997em}. Its extension to NNLO accuracy and the numerical comparison 
with its \emph{iterated} counterpart as discussed above is not pursued in this paper
but will be subject to future work.

Given that the logarithmic contributions to the NS splitting function are subleading up to the NNLL accuracy
considered in this paper, see Ref.~\cite{Kom:2012hd}, no small-$z$ effects have to be considered. 
The usual fixed-order NS evolution equations and kernels should be used instead.

%%%%%%%%%%%%%%%%%%%%%%%%%%%%%%%%%%%%%
\subsection{Numerical Implementation \label{subsec:num}}
%%%%%%%%%%%%%%%%%%%%%%%%%%%%%%%%%%%%%
%
In this section, we will review how to adapt the numerical implementation of the fixed-order results
up to NNLO accuracy, as discussed in Ref.~\cite{Anderle:2015lqa} to include also 
the small-$z$ resummations as discussed above.

%%%%%%%%%%%%%%%%%%%%%%%%%%%%%
% FIGURE 1
%%%%%%%%%%%%%%%%%%%%%%%%%%%%%
\begin{figure}[thb!]
\begin{center}
\includegraphics[width=0.5\textwidth]{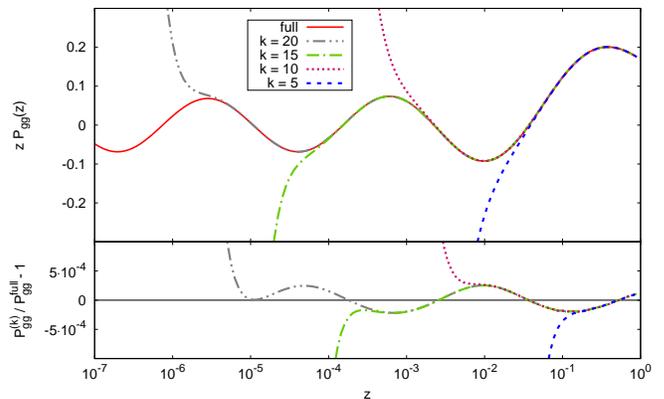}
\end{center}
\vspace*{-0.5cm}
\caption{{\bf Upper panel:} expansion of the splitting function $P_{gg}(z)$ times $z$ at NNLL accuracy
for different upper values of $k$ compared to the fully resummed
expression of Ref.~\cite{Vogt:2011jv, Kom:2012hd}.
{\bf Lower panel:} deviation of the full and ${\cal{O}}(k)$ expanded results. 
All functions are evaluated at $Q^2 = 110\,\mathrm{GeV}^2$ and 
$N_f = 5$ active flavors. \label{fig:Pgg_NNLL_convergence}}
\end{figure}
Following the discussions on the iterated solution in Sec.~\ref{subsec:evol}, we start with 
assessing the order $k$ in $\boldsymbol P^{T\; \text {N}^\kappa\text{LL},(k)}$ 
that is necessary to capture the behavior of fully resummed series down to values
of $z$ relevant for phenomenological studies of SIA data in terms of scale-dependent FFs.
To this end, we study the convergence of the series expansion of the resummed expressions
when evaluated up to a certain order $k$. This is achieved by first expanding the resummed splitting functions 
in Mellin $N$ space and then using an appropriate numerical Mellin inversion, see below, 
to compare the expanded result with the fully resummed splitting functions in $z$-space given in 
\cite{Vogt:2011jv, Kom:2012hd}.
A typical example, the gluon-to-gluon splitting function, is shown in Fig.~\ref{fig:Pgg_NNLL_convergence}.
As can be seen, $k = 20$ in the expansion 
is accurate at a level of less than $0.3 \permil$ differences down to values of $z \approx 10^{-5}$. 
This is more than sufficient for all phenomenological studies as SIA data only extend down to about
$z=10^{-3}$ as we shall discuss later.

However, the splitting functions enter the scale evolution of the FFs in a highly non-trivial way, 
cf.\ Eqs.~\eqref{eq:mtevolution} and \eqref{eq:Rmatrix}, such that this convergence property
does not directly imply that the effects of truncating the expansion at ${\cal{O}}(k=20)$
are also negligible in the solution of the evolution equations. 
To explore this further, we recall that the $N$-space version of Eq.~\eqref{eq:evolution} reads
\begin{equation}
\label{eq:evolution_Nspace}
\frac{\partial}{\partial\ln\mu^2}D^h_i(N,\mu^2)= \sum_j \widetilde{P}^T_{ji}(N,\mu^2)\cdot D^h_j\left(N,\mu^2\right)\, ,
\end{equation}
where $\widetilde{P}^T_{ji}$ is the $ij$-entry of the $2\times 2$ singlet matrix in \eqref{eq:PTtilde}. 
One can solve this equation numerically with the fully resummed kernels, assuming some initial set of FFs, 
and compare the resulting, evolved distributions with the corresponding FFs obtained from the iterative solution of
Eq.~\eqref{eq:mtgeneralsolution} at ${\cal{O}}(k=20)$ defined in Sec.~\ref{subsec:evol}.
Again, we find that the two results agree at a level of a few per mill for $z \gtrsim 10^{-5}$, i.e.,
after transforming the evolved FFs from $N$ to $z$-space.

%%%%%%%%%%%%%%%%%%%%%%%%%%%%%
% FIGURE 2 
%%%%%%%%%%%%%%%%%%%%%%%%%%%%%
\begin{figure}[thb!]
\begin{center}
\includegraphics[width=0.4\textwidth]{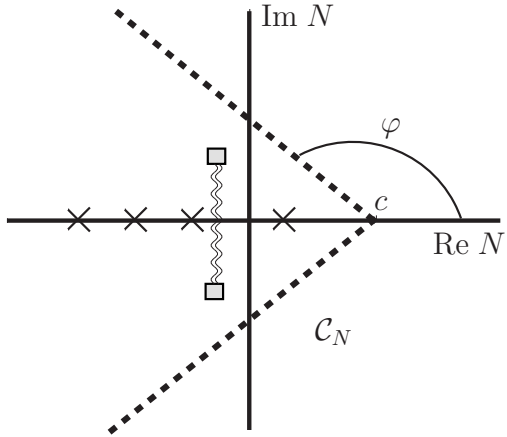}
\end{center}
\vspace*{-0.3cm}
\caption{The dashed line represents the standard contour
$\mathcal{C}_N$ in the complex $N$ plane for the inverse Mellin transformation (\ref{eq:inverse1}).
The poles of the integrand along the real axis are schematically represented by crosses, 
whereas the poles lying in the complex plane away from the real axis are represented by squares. 
The branch cut is illustrated by the wiggly line.
\label{fig:tiltedcontour}
}
\end{figure}
In general, the Mellin inversion of a function $f(N)$ is defined as
\be
\label{eq:Mellin_inversion_general}
f(z) = \f{1}{2\pi i}\int_{{\cal C}_N} dN\, z^{-N}\, f(N)\,,
\ee 
where the contour ${\cal C}_N$ in the complex plane is usually taken parallel to the imaginary axis with
all singularities of the function $f(N)$ to its left. For practical purposes, i.e. faster numerical
convergence, one chooses a deformed contour instead,
which can be parametrized in terms of a real variable $t$, an angle $\varphi$, 
and a real constant $c$ as $N(t) = c + t e^{i\varphi}$; 
see Fig.~\ref{fig:tiltedcontour} for an illustration of the chosen path and Ref.~\cite{ref:pegasus}
for further details.

In order to properly choose the contour parameters $c$ and $\varphi$, we proceed
as in Ref.~\cite{Anderle:2015lqa} and analyze the pole structure of the evolution kernels 
$\mathcal{K}_{ij}^T$. 
They are defined as the entries of the $2\times 2$ time-like 
evolution matrix in 
\be
\label{eq:mtgeneralsolutionschema}
\boldsymbol D^h(N,a_s)= \Bigg (\begin{matrix}\mathcal{K}_{11}^{T}(a_s,a_0,N) & \mathcal{K}^{T}_{12}(a_s,a_0,N)\\[2mm] 
\mathcal{K}^{T}_{21}(a_s,a_0,N) & \mathcal{K}^{T}_{22}(a_s,a_0,N) \end{matrix} \Bigg )\boldsymbol D^h(N,a_0)\,,
\ee
i.e.\ they encompass all the evolution matrices $\boldsymbol U_k$ on the right-hand-side of 
Eq.~(\ref{eq:mtgeneralsolution}).

In complete analogy to what was found in Ref.~\cite{Blumlein:1997em} in the space-like case,
the fully resummed time-like splitting functions exhibit additional singularities as compared to the 
fixed order expressions. Their location in the complex plane away from the real axis depends on the value of $a_s$.
More specifically, if we consider, for instance, $P_{gg}^T$ at NLL \cite{Kom:2012hd}, one can
identify terms proportional to $\left(\sqrt{1 + 32 C_A a_s(\mu) / (N-1)^2}\right)^{-1}$ 
which lead to poles at $N = 1 \pm i \sqrt{32 C_A a_s(\mu)}$ that are connected by a branch cut. 
If we had chosen to directly solve Eq.~\eqref{eq:evolution_Nspace} numerically 
with the fully resummed splitting functions, the appropriate choice of contour for the 
Mellin inversion in Fig.~\ref{fig:tiltedcontour} would have to be $\mu$ dependent 
as the position of these poles, denoted by the squares, depends on $a_s(\mu)$.

%%%%%%%%%%%%%%%%%%%%%%%%%%%%%
% FIGURE 3
%%%%%%%%%%%%%%%%%%%%%%%%%%%%%
\begin{figure}[thb!]
\vspace*{-0.5cm}
\begin{center}
\includegraphics[width=0.48\textwidth]{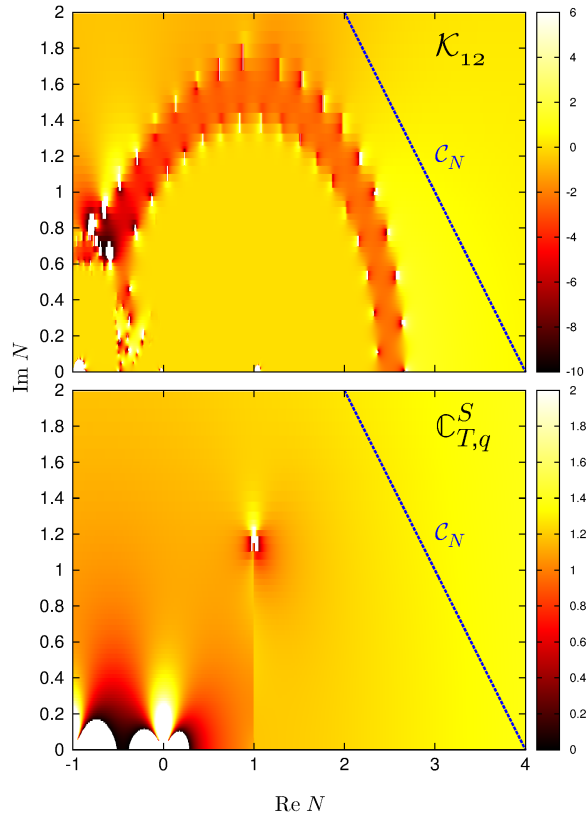}
\end{center}
\vspace*{-0.8cm}
\caption{{\bf Upper panel:} real part of $\mathcal{K}_{12}$ in Eq.~(\ref{eq:mtgeneralsolutionschema}) 
in a portion of the complex $N$ plane. 
{\bf Lower panel:} as above but for the coefficient function $\mathbb{C}^S_{T,q}(N)$. 
Both quantities are computed at NLO+NNLL accuracy for $Q^2=110\,\mathrm{GeV}^2$. 
The line corresponds to the contour ${\cal C}_N$ in \eqref{eq:inverse1}. 
\label{fig:contour} }
\end{figure}
In the iterative solution, which we adopt throughout, only the 
expanded splitting functions $\boldsymbol P^{T\; \text {N}^\kappa\text{LL},(k)}$ 
enter the $\mathcal{K}_{ij}^T$ in Eq.~(\ref{eq:mtgeneralsolutionschema}).
Therefore, the evolution is not affected by the singularities present in the fully resummed kernels,
and a unique, $\mu$-independent choice of the contour parameters $c$ and $\varphi$ is still possible. 
In our numerical code, we take $c = 4$ and $\varphi = 3/4\,\pi$. This choice also tames numerical instabilities 
generated, in particular, by large cancellations caused by the oscillatory behavior in the vicinity
of the $N=1$ pole. This is visualized in the upper panel of Fig.~\ref{fig:contour}. 
Here, we show the real part of the singlet evolution kernel $\mathrm{Re}\{\mathcal{K}_{12}^T\}$ defined
in Eq.~(\ref{eq:mtgeneralsolutionschema}) at NLO+NNLL accuracy and $Q^2=110\,\mathrm{GeV}^2$. 
The numerical instabilities are well recognizable near the $N=1$ pole.
 
Finally, in order to perform a fit of FFs based on SIA data one 
 has to compute the multiplicities as defined in Eq.~\eqref{eq:nnlostructure}. 
As was mentioned above, in order to arrive at a fast but reliable numerical implementation
of the fitting procedure, we choose to evaluate the SIA cross section also in Mellin moment space and, then, perform a 
numerical inverse transformation to $z$-space.
Schematically, one has to compute integrals of the form
\be\label{eq:inverse1}
D(z) \otimes \mathbb{C}(z)=\f{1}{2\pi i}\int_{{\cal C}_N} dN\, z^{-N}\, D(N)\,\mathbb{C}\left(N\right) \, ,
\ee
where the FFs $D(N)$ are given by Eq.~\eqref{eq:mtgeneralsolution}; for brevity, we 
have omitted any dependence on the scale $\mu$ and the parton flavor. 
In principle, while performing the Mellin inversion, one has to 
deal with the same kind of \mbox{$a_s$-dependent} singularities in the fully resummed
resummed coefficient functions, cf.\ Ref.~\cite{Kom:2012hd},
that we have just encountered in the resummed splitting functions. 
In the lower panel of Fig.~\ref{fig:contour}, we show the real part of the 
coefficient function $\mathbb{C}^S_{T,q}(N)$ for which the pole structure and the branch cut 
are again well recognizable. 
However, for the typical scales relevant for a phenomenological analysis 
($\mu = 10.5-91.2\,\mathrm{GeV}$; see Sec.~\ref{sec:fit}), our choice of contour $\mathcal{C}_N$ 
is nevertheless applicable since the position of the singularities does not change
considerably in this range of energies.

%%%%%%%%%%%%%%%%%%%%%%%%%%%%%%%%%%%%%%%%%%%%%%%%%%%%%%%
\section{Phenomenological Applications \label{sec:fit}}
%%%%%%%%%%%%%%%%%%%%%%%%%%%%%%%%%%%%%%%%%%%%%%%%%%%%%%%
%
In the literature, small-$z$ resummations have been exploited to exclusively study the
fixed $N=1$ moment of integrated hadron multiplicities in SIA, in particular, their
scale evolution and the shift of the peak of the multiplicity distribution with energy \cite{ref:n1moment_pheno}.
In this section, we will extend these studies to the entire $z$-range and 
present a first phenomenological analysis of SIA data with identified pions 
in terms of FFs up to NNLO+NNLL accuracy. 
More specifically, we use the same data sets as in a recent fixed-order 
fit of parton-to-pion FFs at NNLO accuracy \cite{Anderle:2015lqa}.
In Sec.~\ref{subsec:fit} we perform various fits to SIA data with and without
making use of small-$z$ resummations to quantify their phenomenological relevance.
The impact of small-$z$ resummations on the residual dependence on the factorization scale 
is studied in Sec.~\ref{subsec:appl}.
%
%%%%%%%%%%%%%%%%%%%%%%%%%%%%%%%%%%%%%
\subsection{Fits to SIA data and the relevance of resummations \label{subsec:fit}}
%%%%%%%%%%%%%%%%%%%%%%%%%%%%%%%%%%%%%
%
To set up the framework for fitting SIA data with identified pions, we closely follow
the procedures outlined in Refs.~\cite{ref:dss,ref:dss2,Epele:2012vg,ref:dssnew,Anderle:2015lqa}.
Thus, we adopt the same flexible functional form
\be\label{eq:Dparam}
D_i^{\pi^{+}}(z,\mu_0^2)=\f{N_i\, z^{\alpha_i}(1-z)^{\beta_i}[1+\gamma_i(1-z)^{\delta_i}]}
{B[2+\alpha_i,\beta_i+1]+\gamma_i B[2+\alpha_i,\beta_i+\delta_i+1]} 
\ee
to parametrize the non-perturbative FFs for charged pions
at some initial scale $\mu_0$ in the commonly adopted $\overline{\mathrm{MS}}$ scheme. 
Other than in Refs.~\cite{ref:dss,ref:dss2,Epele:2012vg,ref:dssnew,Anderle:2015lqa}, we choose, however,
\mbox{$\mu_0=10.54$ GeV}, which is equivalent to the lowest c.m.s. energy $\sqrt{S}$ of the the data sets relevant for the fit.
This choice is made to avoid any potential bias in our comparison of fixed-order and resummed
extractions of FFs from starting the scale evolution at some lowish, hadronic scale ${\cal{O}}(1\,\mathrm{GeV})$
where non-perturbative corrections, i.e., power corrections, might be still of some relevance.
The Euler Beta function $B[a,b]$ in the denominator of (\ref{eq:Dparam})
is introduced to normalize the parameter $N_i$ for each flavor $i$ to its contribution 
to the energy-momentum sum rule.
 
As can be inferred from Eq.~(\ref{eq:nnlostructure}), SIA is only sensitive to 
certain combinations of FFs, namely the sum of quarks and anti-quarks, $q_i+\bar q_i$, 
for a given flavor $i$ and the gluon $D_g^h$. 
Therefore, in all our fits, we only consider FFs for these flavor combinations, i.e., 
$u+\bar u$, $d+\bar d$, $s+\bar s$, $c+\bar c$, $b+\bar b$, and $g$, each parametrized
by the ansatz in (\ref{eq:Dparam}).
The treatment of heavy flavor FFs, i.e., charm and bottom quark and antiquark, proceeds in
the same, non-perturbative input scheme (NPIS) used in Ref.~\cite{Anderle:2015lqa} and in 
the global analyses of \cite{ref:dss,ref:dss2,Epele:2012vg,ref:dssnew}. 
More specifically, non-perturbative input distributions $D_{c+\bar{c},b+\bar{b}}^h(z,m_{c,b}^2)$,
are introduced as soon as the scale in the evolution 
crosses the value of the heavy quark pole mass $m_{c,b}$, for which we use 
$m_c=1.4\,\mathrm{GeV}$ and $m_b=4.75\,\mathrm{GeV}$, respectively.
At the same time, the number of active flavors is increased by one, $N_f \to N_f+1$, in all expressions
each time a flavor threshold is crossed. 
Since we use \mbox{$\mu_0 = 10.54$ GeV $> m_b$}, this never actually happens in the present fit.
The parameters of $D_{c+\bar{c},b+\bar{b}}^h(z,m_{c,b}^2)$
are determined by the fit to data according to the Eq.~(\ref{eq:Dparam}).
We note that a general-mass variable flavor number scheme for treating the heavy quark-to-light hadron
FFs has been recently put forward in Ref.~\cite{Epele:2016gup}. Since this scheme, as well as other
matching prescriptions \cite{Cacciari:2005ry}, are only available up to NLO accuracy, we refrain from 
using them in our phenomenological analyses. 

Rather than fitting the initial value of the strong coupling at some reference scale in order 
to solve the RGE governing its running, we adopt the following boundary conditions
$\alpha_s(M_Z)=0.135$, $0.120$, and $0.118$
at LO, NLO, and NNLO accuracy, respectively, from the recent
MMHT global analysis of PDFs \cite{Harland-Lang:2014zoa}. When we turn on small-$z$ resummations
at a given logarithmic order N$^{m}$LL in our fit, we keep the $\alpha_s$ value as appropriate for
the underlying, fixed-order calculation to which the resummed results are matched. For instance, in a fit
at NLO+NNLL accuracy, we use the $\alpha_s$ value at NLO.

In the present paper, we are mainly interested in a comparison of fixed-order fits with corresponding
analyses including small-$z$ resummations to determine the phenomenological impact of the latter. 
We make the following selection of data to be included in our fits. 
First of all, as in Ref.~\cite{Anderle:2015lqa}, we limit ourselves to SIA with identified pions since these data
are the most precise ones available so far. They span a c.m.s.\ energy range from 
$\sqrt{S}\simeq 10.5\,\mathrm{GeV}$ at the $b$-factories at SLAC and KEK 
to $\sqrt{S}=M_Z\simeq 91.2\,\mathrm{GeV}$ at the CERN-LEP.
The second, more important selection cut concerns the lower value in $z$ accepted in the fit.
Traditionally, fits of FFs introduce a minimum value $z_{\min}$ of the energy fraction $z$ 
in the analyses below which all SIA data are discarded and FFs should not be used in other processes. 
This rather ad hoc cut is mainly motivated by kinematic considerations, more specifically, by the finite hadron mass 
or other power corrections which are neglected in the factorized framework \cite{ref:nason}.
Hadron mass effects in SIA have been investigated to some extent in~\cite{ref:hmc+res} but there is no systematic way to
properly include them in a general process~\cite{Christova:2016hgd}, i.e., ultimately in a global analysis of FFs.
In case of pion FFs, one usually sets $z_{\min}=0.1$ \cite{ref:dss,ref:dssnew} 
or $z_\text{min}=0.075$ \cite{Anderle:2015lqa}.

\begin{figure}[t!]
\begin{center}
\includegraphics[width=0.5\textwidth]{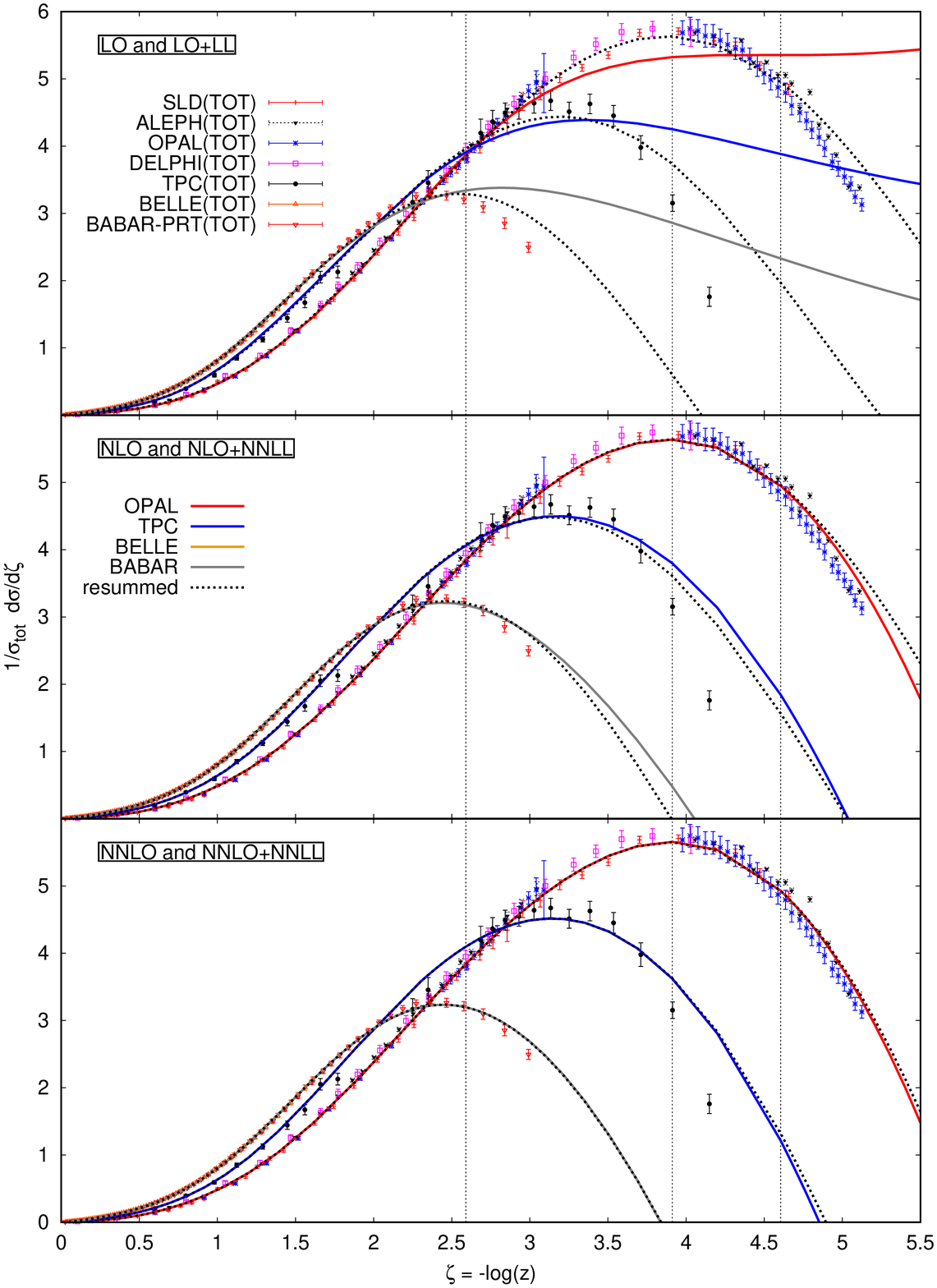}
\end{center}
\caption{Pion multiplicity data  \cite{ref:belledata,ref:babardata,ref:tpcdata,ref:slddata,ref:alephdata,ref:delphidata,ref:opaldata}
 included in the analyses as a function of $\zeta=\log{(1/z)}$
compared to the results of various fits without (solid lines) and with (dotted lines) 
small-$z$ resummations. All curves refer to the central choice of scale $\mu=Q$.
The top, middle, and lower panel shows the results at LO and LO+LL, 
NLO and NLO+NNLL, and NNLO and NNLO+NNLL accuracy, respectively.
The vertical dotted lines illustrate, from left to right, the 
lower cuts $z_{\min}=0.075$ adopted in \cite{Anderle:2015lqa}, and
$z_{\min}=0.02$ and $0.01$ used in all our fits for the TPC data and otherwise, respectively. 
\label{fig:data_vs_theory}
}
\end{figure}

The two main assets one expects from small-$z$ resummations, and which we want to
investigate, are an improved scale dependence and an extended range towards lower values 
of $z$ in which data can be successfully described. 
For this reason, we have systematically explored to which extent one can lower the cut $z_{\min}$ in a fit
to SIA data once resummations as outlined in Sec.~\ref{sec:nnlo} are included.
It turns out, that for the LEP data, taken at the highest c.m.s.\ energy of $\sqrt{S}=91.2$ GeV, 
we can extend the $z$-range of our analyses from $0.075<z<0.95$ used in the NNLO fit \cite{Anderle:2015lqa}
to $0.01<z<0.95$. 
Unfortunately, any further extension to even lower values of $z$ is hampered by the fact that
two of the data sets from LEP, the ones from ALEPH \cite{ref:alephdata} and OPAL \cite{ref:opaldata}, 
appear to be mutually inconsistent below $z\simeq 0.01$, see Fig.~\ref{fig:data_vs_theory}. Including these
data at lower $z$, always lets the fits, i.e., the minimization in the
multi-dimensional parameter space defined by Eq.~(\ref{eq:Dparam}), go astray and the convergence is very poor.

For the relevant data sets at lower c.m.s.\ energies,
TPC \cite{ref:tpcdata} ($\sqrt{S}=29$ GeV), BELLE \cite{ref:belledata} ($\sqrt{S}=10.52$ GeV), 
and BABAR \cite{ref:babardata} ($\sqrt{S}=10.54$ GeV),
the above mentioned problems related to the finite hadron mass arise at small values of $z$. 
A straightforward, often used criterion to assess the relevance of hadron mass effects is to compare
the scaling variable $z$, i.e. the hadron's energy fraction $z=2\,E_h/Q$ in a c.m.s.\ frame, 
with the corresponding three-momentum fraction $x_p$ which is often used in experiments.
Since they are related by $x_p=z-2m_h^2/(z Q^2) + {\cal{O}}(1/Q^4)$ \cite{ref:nason}, i.e.,
they coincide in the massless limit, any deviation
of the two variables gives a measure of potentially important power corrections.
To determine the cut $z_{\min}$ for a given data set,
we demand that $z$ and $x_p$ are numerically similar at a level of 10 to at most $15\%$.
The BELLE data are limited to the range $z>0.2$ \cite{ref:belledata}, where $z$ and $x_p$ differ by less than $1\%$.
BABAR data are available for $z\gtrsim 0.05$, which translates in a maximum difference 
of the two variables of about $14\%$. Concerning the TPC data, we had to place a lower cut
$z_\text{min}=0.02$ to arrive at a converged fit,
which corresponds to a difference of approximately $11\%$ between $z$ and $x_p$.
After imposing these cuts, the total amount of data points taken into account in our fits is 436.
We note that, in general, the interplay between small-$z$ resummations and the various sources of power corrections 
poses a highly non-trivial problem which deserves to be studied further in some dedicated future work. 
%
%%%%%%%%%%%%%%%%%%%%%%%%%%%%%
% FIGURE 4
%%%%%%%%%%%%%%%%%%%%%%%%%%%%%
%%%%%%%%%%%%%%%%%%%%%%%%%%%%%

It is also worth mentioning that with the lowered kinematic cut $z_\text{min}$,
we achieve a better convergence of our fits with our choice of a larger initial scale \mbox{$\mu_0=10.54$ GeV}
in Eq.~(\ref{eq:Dparam}).
Starting the scale evolution from a lower value $\mu_0={\cal{O}}(1)\,\mathrm{GeV}$, 
like in the NNLO analysis of Ref.~\cite{Anderle:2015lqa},
leads, in general, to less satisfactory fits in terms of their total $\chi^2$ value which is used to
judge the quality of the fits.  
This could relate to the fact that other types of power corrections have to be considered as well
when evolving from such a low energy scale in order to be able to describe the shape of the 
differential pion multiplicities, cf.\ Fig.~\ref{fig:data_vs_theory}, measured in experiment. 
To corroborate this hypothesis is well beyond the scope of this paper.
In any case, our choice of $\mu_0$ is certainly in a region where the standard perturbative framework 
can be safely applied and meaningful conclusions on the impact of small-$z$ resummations in SIA 
can be drawn. 
 
Turning back to the choice of our flexible ansatz for the FFs, it is well known that fits based solely
on SIA data are not able to constrain all of the free parameters in Eq.~(\ref{eq:Dparam}) for each of
the flavors $i$. As was shown in the global analysis of SIA, SIDIS, and $pp$ data in \cite{ref:dssnew},
charge conjugation and isospin symmetry are well satisfied for pions. Therefore, we impose
the constraint $D^{\pi^\pm}_{u+\bar u}=D^{\pi^\pm}_{d+\bar d}$.
We further limit the parameter space associated with the large-$z$ region 
by setting $\delta_{g,s+\bar{s},c+\bar{c}} = 0$ and $\gamma_{g,s+\bar{s},c+\bar{c}} = 0$. 
Note that in contrast to Ref.~\cite{Anderle:2015lqa}, we are now able to keep $\beta_g$ 
as a free parameter in the fits.

The remaining 19 free parameters are then determined by a standard $\chi^2$ minimization 
procedure as described, for example, in Ref.~\cite{ref:dssnew}. 
The optimal normalization shifts for each data set are computed analytically. They
contribute to the total $\chi^2$ according to the quoted experimental normalization uncertainties;
see, e.g., Eq.~(5) in Ref.~\cite{ref:dssnew} for further details. 
The resulting $\chi^2$-values, the corresponding ``penalties'' from the
normalization shifts, and the $\chi^2$ per degree of freedom (dof) are listed in Tab.~\ref{tab:fitres}
for a variety of fits with a central choice of scale $\mu=Q$.
Results are given both for fits at fixed order (LO, NLO, and NNLO) accuracy 
and for selected corresponding fits obtained with small-$z$ resummations. Here, all cross sections 
are always matched to the fixed order results according to the procedures described in 
Sec.~\ref{subsec:scales} and Sec.~\ref{subsec:evol}.
More specifically, we choose the logarithmic order in such a way that
we do not resum logarithmic contributions which are not present in the fixed-order result. 
For this reason, we match the LO calculation only with the LL resummation as the only logarithmic 
contribution at LO is of LL accuracy; cf.\ Tabs.~\ref{tab:Ndependence_coefs} and
\ref{tab:Ndependence_splittings}. Using the same reasoning, we match NLO with the NNLL resummed results. 
Finally, at NNLO accuracy five towers of small-$z$ logarithms are present. However, 
the most accurate resummed result currently available is at NNLL accuracy which includes the first
three towers. Thus, we can match NNLO only with NNLL.
%
%%%%%%%%%%%%%%%%%%%
% TABLE II
%%%%%%%%%%%%%%%%%%
\begin{table}[th!]
\caption{\label{tab:exppiontab} The obtained $\chi^2$-values, the 
``penalties'' from normalization shifts, and the $\chi^2/\mathrm{dof}$ 
for the fits at fixed order and resummed accuracy as described in the text. }
\begin{ruledtabular}
\begin{tabular}{lccc}
\label{tab:fitres}
accuracy & $\chi^2$ & norm shift & $\chi^2/\mathrm{dof}$\\ \hline
LO & 1260.78 & 29.02 & 2.89\\
NLO & 354.10 & 10.93 & 0.81\\
NNLO & 330.08 & 8.87 & 0.76\\
LO+LL & 405.54 & 9.83 & 0.93\\
NLO+NNLL & 352.28 & 11.27 & 0.81\\
NNLO+NNLL & 329.96 & 8.77 & 0.76\\
\end{tabular}
\end{ruledtabular}
\end{table}
It should be stressed that the results for the fixed-order fits are not directly comparable 
to the ones given in Ref.~\cite{Anderle:2015lqa} since we use more data
points at lower values of $z$, a slightly different set of fit parameters, and a different
initial scale $\mu_0$.
However, the main aspects of these fits remain the same and can be read off directly 
from Tab.~\ref{tab:fitres}: a LO fit is not able to describe
the experimental results adequately. The NLO fit already gives an acceptable result, 
which is further improved upon including NNLO corrections.
Compared to the corresponding fixed-order results, the fits including also all-order resummations of small-$z$ 
logarithms exhibit, perhaps somewhat surprisingly, only a slightly better total $\chi^2$, 
except for the LO+LL fit, where resummation leads to a significant improvement in its quality.
The small differences in $\chi^2$ between fits at NNLO and NNLO+NNLL accuracy 
are not significant. 
Hence, we must conclude that in the $z$-range covered by the experimental results, 
NNLO expressions already capture most of the relevant features to yield a satisfactory fit
to the SIA data with identified pions.

The same conclusions can be reached from Fig.~\ref{fig:data_vs_theory}, where we 
compare the used inclusive pion multiplicity data in SIA with the 
theoretical cross sections at different levels of fixed- and logarithmic-order 
obtained from the fits listed in Tab.~\ref{tab:fitres}. The theoretical curves 
are corrected for the optimum normalization shifts computed for each set of data.
For the sake of readability, we only show a single curve for the different 
experiments at $\sqrt{S}=M_Z$ which is corrected for the normalization shift obtained for
the OPAL data. The individual normalization shifts for the other sets
are, however, quite similar. 
We refrain from showing the less precise flavor-tagged data which are, nevertheless, 
also part of the fit.
The vertical dotted lines in Fig.~\ref{fig:data_vs_theory}
indicate the lower cuts in $z$ applied for the data sets at different c.m.s.\ energies as
discussed above. The leftmost line (corresponding to $z_\text{min}=0.075$) 
is the cut used in the NNLO analysis in Ref.~\cite{Anderle:2015lqa}. 
Both, the data and the calculated multiplicities are shown as
a function of $\zeta\equiv -\log z$. 

\begin{figure}[thb!]
\vspace*{-0.3cm}
\begin{center}
\includegraphics[width=0.5\textwidth]{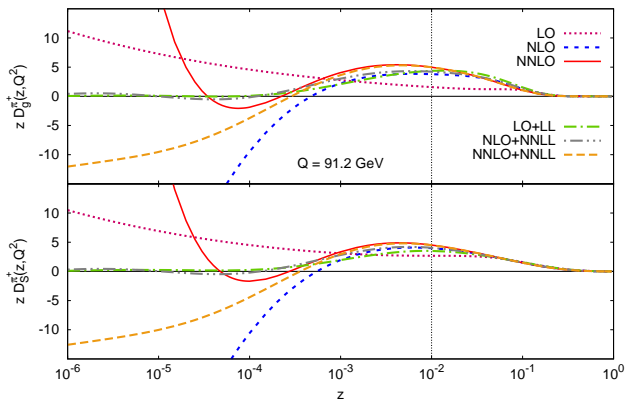}
\end{center}
\vspace*{-0.8cm}
\caption{$z$ times the obtained gluon (upper panel) and singlet (lower panel) FFs as a function of $z$,
evaluated at $Q=91.2\;\mathrm{GeV}$ for the different fits listed in Tab.~\ref{tab:fitres}. 
The singlet is shown for $N_f=5$ active flavors. The fitted $z$-range, $z>0.01$, is to the right
of the dotted vertical line.
\label{fig:Dg_Dsinglet}
}
\end{figure}

In Fig.~\ref{fig:Dg_Dsinglet}, we plot $z$ times the gluon and singlet FFs
for positively charged pions, $D_g^{\pi^+}(z,Q^2)$
and $D_{S}^{\pi^+}(z,Q^2)$, respectively,
resulting from our fits given in Tab.~\ref{tab:fitres}.
The FFs are computed at $Q=M_Z=91.2\,\mathrm{GeV}$ and in a range of $z$ shown extending well below
the $z_{\min}=0.01$ cut above which they are constrained by data.
We would like to point out that the resummed (and matched) results for which we have full control 
over all logarithmic powers (i.e.\ for LO+LL and NLO+NNLL) are well behaved at small-$z$ 
and show the expected oscillatory behavior with $z$
which they inherit from the resummed splitting functions through evolution. 
The latter behave like different combinations of Bessel functions when the Mellin inverse back
to $z$-space is taken;
for more details see Ref.~\cite{Kom:2012hd}. 
The singlet and gluon FFs at NNLO+NNLL accuracy still diverge for $z\to 0$
(i.e.\ they  turn to large negative values in the $z$-range shown in Fig.~\ref{fig:Dg_Dsinglet})  
since we do not have control over all five logarithmic powers  
that appear in a fixed-order result at NNLO; cf.\ Tabs.~\ref{tab:Ndependence_coefs} and
\ref{tab:Ndependence_splittings}.
However, the resummation of the three leading towers of logarithms, 
considerably tames the small-$z$ singularities as compared to the corresponding result obtained at NNLO.
%
%%%%%%%%%%%%%%%%%%%%%%%%%%%%%
% FIGURE 4
%%%%%%%%%%%%%%%%%%%%%%%%%%%%%

\begin{figure}[thb!]
\begin{center}
\includegraphics[width=0.5\textwidth]{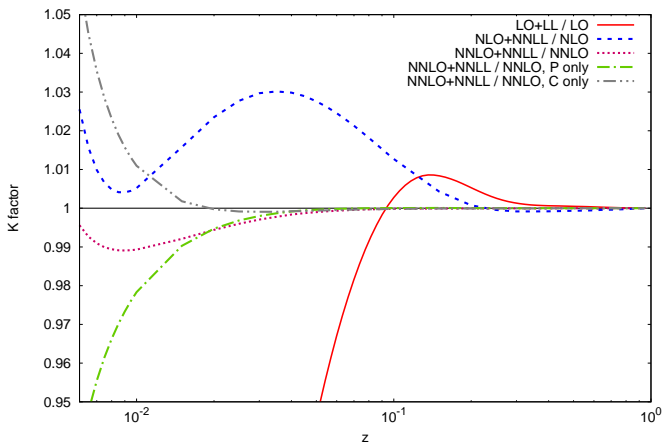}
\end{center}
\caption{$K$-factors as defined in Eq.~(\ref{eq:defK})
at LO+LL, NLO+NNLL, and NNLO+NNLL accuracy at $Q=91.2$ GeV in the range of $z$ relevant
for phenomenology. 
In addition, we show NNLO+NNLL results where the resummations are only performed either
for the coefficient functions (''$C$ only'') or for the splitting functions (''$P$ only'').
\label{fig:Kfactors_z}
}
\end{figure}

Finally, to further quantify the impact of small-$z$ resummations in the range of $z$ relevant
for phenomenology, Fig.~\ref{fig:Kfactors_z} shows 
the $K$-factors at scale $Q=91.2\,\mathrm{GeV}$
for the pion multiplicities (\ref{eq:nnlostructure}) obtained in our fits. 
Schematically, they are defined as
\be
\label{eq:defK}
K \equiv \frac{\mathbb{C}^\text{FO + Res} \otimes D^\text{FO + Res}}{\mathbb{C}^\text{FO} \otimes D^\text{FO}}\,.
\ee
Here, $\mathbb{C}^\text{FO}$ and $\mathbb{C}^\text{FO+Res}$ denote the fixed-order coefficient functions 
at LO, NLO, and NNLO accuracy and the corresponding resummed and matched coefficient functions, respectively.
Likewise, $D^\text{FO}$ and $D^\text{FO+Res}$ are the FFs evolved with splitting functions 
at fixed order and resummed, matched accuracy, respectively. 
In order to assess the relevance of the small-$z$ resummations independent of the details of the non-perturbative
input for the FFs at scale $\mu_0$, we adopt the same FFs for both calculating the numerator and the denominator. 
In each computation of $K$, we select the set of FFs obtained from the corresponding fixed-order fit 
and the different logarithmic orders of the resummations are chosen as discussed and given in 
Tab.~\ref{tab:exppiontab}.

%%%%%%%%%%%%%%%%%%%%%%%%%%%%%
% FIGURE 6
%%%%%%%%%%%%%%%%%%%%%%%%%%%%%
By comparing the results for the $K$-factors at LO+LL, NLO+NNLL, and NNLO+NNLL accuracy, 
it can be infered that the corrections due to the small-$z$ resummations start to become appreciable
at a level of a few percent already below $z\simeq 0.1$. 
As one might expect, resummations are gradually less important when the perturbative accuracy 
of the corresponding fixed-order baseline is increased, i.e., 
the NNLO result already captures most of the small-$z$ dynamics relevant for phenomenology whereas
the differences between LO and LO+LL are still sizable. This explains the pattern of $\chi^2$ values
we have observed in Tab.~\ref{tab:exppiontab}.
In addition, Fig.~\ref{fig:Kfactors_z} also gives the $K$-factor at NNLO+NNLL accuracy 
where the small-$z$ resummations are only performed either
for the coefficient functions (labeled as ''$C$ only'') {\em or} for the splitting functions (''$P$ only'').
By comparing these results with the full $K$-factor at NNLO+NNLL accuracy, one can easily 
notice, that there are very large cancellations among the two.

%%%%%%%%%%%%%%%%%%%%%%%%%%%%%%%%%%%%%
\subsection{Scale dependence \label{subsec:appl}}
%%%%%%%%%%%%%%%%%%%%%%%%%%%%%%%%%%%%%
%
In this section, the remaining scale dependence of the resummed expressions is studied
and compared to the corresponding fixed-order results. The
scale-dependent terms are implemented according to the discussions in Sec.~\ref{subsec:scales}.
As usual, we use the iterated solution with up to $n=20$ terms in the perturbative expansion.

As was already observed in the NNLO analysis of Ref.~\cite{Anderle:2015lqa}, the dependence on the 
factorization scale $\mu_F$ in SIA is gradually reduced the more higher order corrections are considered 
in the perturbative expansion. This is in line with the expectation that all artificial scales,
$\mu_F$ and $\mu_R$, should cancel in an all-order result, i.e.~if the series is truncated at
order $m$, the remaining dependence on, say, $\mu_F$ should be of order $a_s^{m+1}$.
Following this reasoning, we do expect a further reduction of the scale dependence 
upon including small-$z$ resummations on top of a given fixed-order calculation;
see Sec.~\ref{subsec:scales}.

Usually, the scale dependence is studied by varying the scale $\mu_F$ by a factor of two or four
around its default (central) value, $\mu_F = Q$ in case of SIA. 
Therefore, we introduce the parameter $\xi \equiv \mu_F^2 / Q^2$; note that in this paper we
keep $\mu_F=\mu_R$ as is commonly done. 
Hence, $\xi=1$ corresponds to the standard choice of scale $\mu_F=Q$.
The conventional way of showing the dependence of a quantity $T$,
like the pion multiplicity (\ref{eq:nnlostructure}), on $\xi$
is to plot the ratio $T(\xi) / T(\xi=1)$ for various values of $\xi$;
in our analyses, we will use $\xi=2$ and $\xi=0.5$.

However, we find that the oscillatory behavior of the resummed splitting and coefficient functions 
causes the SIA multiplicities to become an oscillatory function as well, 
which for certain small values of $z$, well below the cut $z_{\min}$
down to which we fit FFs to data, eventually becomes negative.
Therefore, it is not feasible to utilize the common ratio plots to investigate the resummed scale dependence.
Instead, we decide to study the {\em width} of the scale variation $\Delta_T$ for a quantity $T$, defined as
\ba
\label{eq:width-delta}
\Delta_T(z) &\equiv& \max[T_{\xi=1}(z),T_{\xi=2}(z),T_{\xi=0.5}(z)] \nn \\
 &-& \min[T_{\xi=1}(z),T_{\xi=2}(z),T_{\xi=0.5}(z)] 
\ea
in the range $\xi=[0.5,2]$ as a measure of the residual dependence on $\mu_F$.

%%%%%%%%%%%%%%%%%%%%%%%%%%%%%
% FIGURE 7
%%%%%%%%%%%%%%%%%%%%%%%%%%%%%
\begin{figure}[!ht]
\begin{center}
\includegraphics[width=0.5\textwidth]{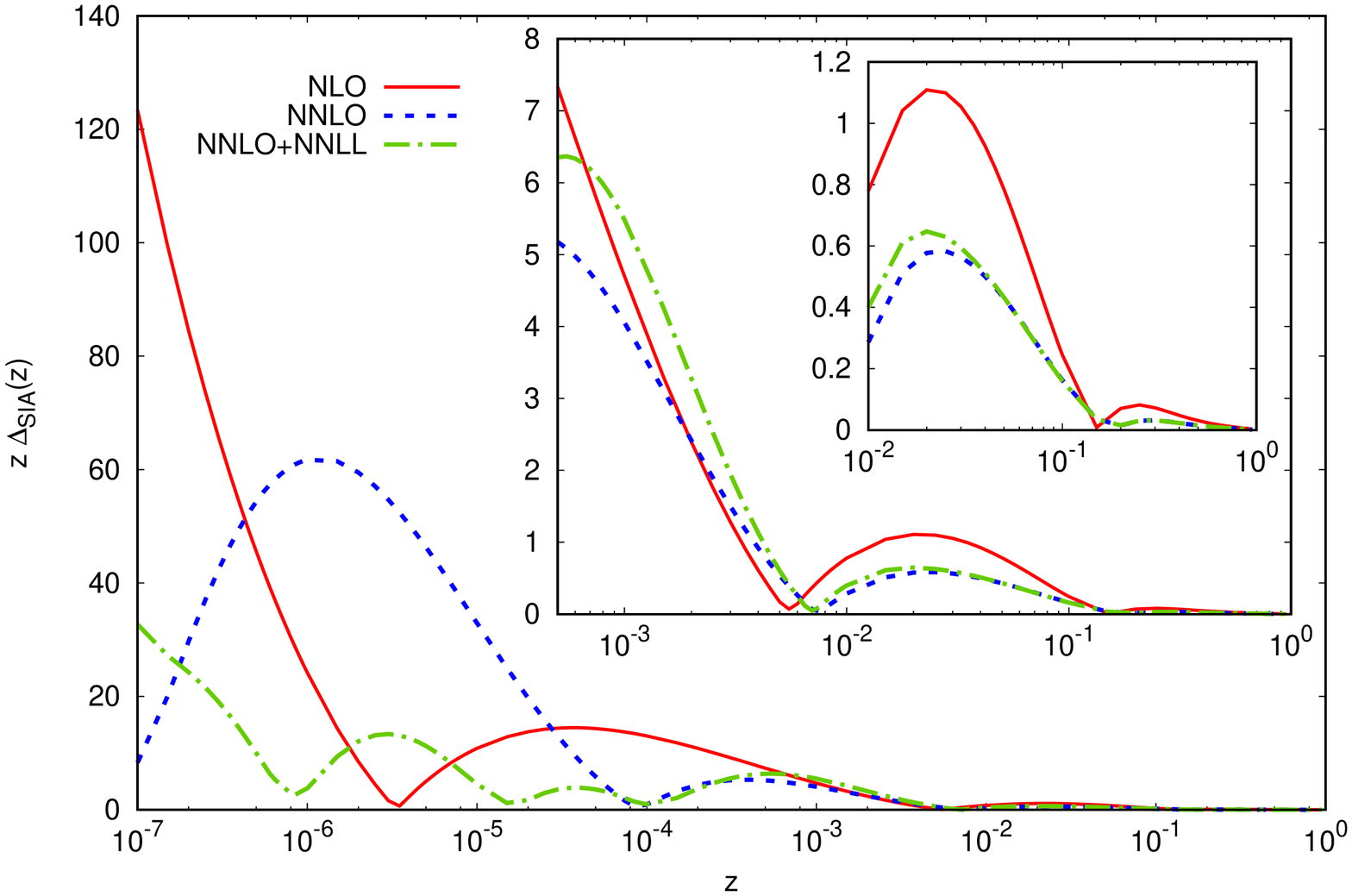}
\end{center}
\caption{$z$ times the width of the scale band $\Delta_{\text{SIA}}$ defined in (\ref{eq:width-delta}) for
for three different ranges of $z$ at NLO, NNLO and NNLO+NNLL accuracy.
All results for the SIA pion multiplicities are obtained for $Q=10.54\,\mathrm{GeV}$; see text.  
\label{fig:SCALE}
}
\end{figure}
%%%%%%%%%%%%%%%%%%%%%%%%%%%%%%
%
In Fig.~\ref{fig:SCALE}, we show $\Delta_{\text{SIA}}(z)$ for the pion multiplicities
\eqref{eq:nnlostructure} at $Q=10.54\,\mathrm{GeV}$
for the two fixed-order fits (NLO and NNLO accuracy) as well as for 
resummed and matched fit at NNLO+NNLL. The main plot, which covers the $z$-range down to $10^{-7}$,
clearly demonstrates that the band $\Delta_{\text{SIA}}$ is, on average, considerably more narrow for
the NNLO+NNLL resummed cross section than for the fixed-order results, according to the expection.
From the middle inset in Fig.~\ref{fig:SCALE}, which shows $z$ values 
relevant for experiments, i.e.\ $z\gtrsim 10^{-3}$, one can infer that the band $\Delta_{\text{SIA}}$ is roughly of the same
size for all calculations and resummations do not lead to any improvement in the scale dependence
in this range. The small inset zooms into the range $z>0.01$, where a similar conclusion can be reached. 

In order to fully understand this behavior, one perhaps would have to include 
the yet missing N$^4$LL corrections, which would allow one to resum all five logarithmic towers present
at NNLO accuracy. The observed result might be due to these missing subleading terms or 
it could be related to some intricate details in the structure of the perturbative series 
in the time-like case at small-$z$.

In any case, one can safely conclude that in the $z$-region relevant 
for phenomenology of SIA, the residual scale dependence of the resummed result does not differ
from the fixed order calculation at NNLO accuracy. The latter is therefore entirely sufficient
for extractions of FFs from SIA data as resummations neither improve the quality of the fit, cf.
Sec.~\ref{subsec:fit} nor do they reduce theoretical uncertainties.
Nonetheless, it important to demonstrate from a theoretical point of view 
that, on average, resummation does achieve smaller scale uncertainties, although 
for values of $z$ that are well outside the range of currently available data.
It should be also kept in mind that the study of the $N=1$ moment of multiplicities,
though not studied in this paper, would not be possible without invoking small-$z$ resummations
as fixed-order results are singular.

%%%%%%%%%%%%%%%%%%%%%%%%%%%%%%%%%%%%%
\section{Conclusions and Outlook \label{sec:conclusions}}
%%%%%%%%%%%%%%%%%%%%%%%%%%%%%%%%%%%%%
%
We have presented a detailed phenomenological analysis of 
small-$z$ resummations in semi-inclusive annihilation, the
time-like scale evolution of fragmentation functions, and their
determination from data.

After detailing the systematics of the enhanced contributions at small
momentum fractions of the observed hadron for both coefficient and 
splitting functions, we have reviewed how to resum
them to all orders in perturbation theory 
up to next-to-next-to-leading logarithmic accuracy. The approach used
in this paper was proposed in the literature and is
based on general considerations concerning all-order mass factorization.
Our results agree with those presented in the literature, and we have
extended them to allow for variations in the factorization and renormalization
scales away from their default values.

Next, we have shown how to properly implement the resummed expressions in 
Mellin moment space and how to set up a solution to the coupled, matrix-valued
singlet evolution equations. The non-singlet sector is subleading and not affected 
by the presently available logarithmic order. For all practical purposes we
advocate an iterated solution for the scale evolution of fragmentation functions,
and we have shown that keeping twenty terms in the expansion of the resummed
expressions is sufficient for all applications. We have also discussed how to
match the resummed towers of logarithms for both the coefficient and 
the evolution kernels to the known fixed-order expressions.
Numerical subtleties in complex Mellin moment space related to finding a proper
choice of contour for the inverse transformation
despite the more complicated structure of singularities 
of the resummed evolution kernels and coefficient functions have been addressed as well.

In the second part of the paper, a first analysis of semi-inclusive annihilation
data with an identified pion in terms of parton-to-pion fragmentation functions
and in the presence of resummations was presented. 
To this end, various fits at different fixed-orders in perturbation theory and levels of
small-$z$ resummations were compared in order to study and quantify the 
phenomenological impact of the latter. 
It turned out that for both the quality of the fit to data and the
reduction of theoretical uncertainties due to the choice of the factorization 
scale, resummations provide only litte improvements with respect to an analysis
performed at fixed, next-to-next-to-leading order accuracy.
At values of the hadron's momentum well outside the range of phenomenological interest,
we did observe, however, a significant improvement in the scale dependence of the 
inclusive pion cross section in the presence of resummations. 
 
Possible future applications of resummations comprise revisiting the analyses
of the first moment of hadron multiplicities available in the literature. 
Here, resummations are indispensable for obtaining a finite theoretical result. 
So far, the main focus was on the energy dependence of the peak of the multiplicity
distribution, its width, and a determination of the strong coupling.
It might be a valuable exercise to merge the available data on the first moment 
and the relevant theoretical formalism with the extraction
of the full momentum dependence of fragmentation functions as described in this paper
to further our knowledge of the non-perturbative hadronization process.

As was pointed out in the paper, a better understanding of the interplay of resummations 
and other sources of potentially large corrections in the region of small momentum fractions
is another important avenue of future studies for time-like processes.
One if not the most important source of power corrections is the hadron mass, which
is neglected in the factorized framework adopted for any analysis of fragmentation functions. 
At variance with the phenomenology of parton distributions functions, where one can access and
theoretically describe the physics of very small momentum fractions, hadron mass corrections prevent
that in the time-like case. In fact, they become an inevitable part and severely 
restrict the range of applicability of fragmentation functions and the theoretical tools
such as resummations.
In addition, resummations can and have been studied for large fractions of the hadron's momentum.
With more and more precise data becoming available in this kinematical regime, it would be 
very valuable to incorporate also these type of large logarithms into the analysis framework for
fragmentation functions at some point in the future.

%%%%%%%%%%%%%%%%
\section*{Acknowledgments}
%%%%%%%%%%%%%%%%
%
We are grateful to W.\ Vogelsang and A.\ Vogt
for helpful discussions and comments.
D.P.A.\ acknowledges partial support from the 
Fondazione Cassa Rurale di Trento.
D.P.A.\ was supported by the Deutsche Forschungsgemeinschaft (DFG) under 
grant no. VO 1049/1, T.K.\ by
the Bundesministerium f\"{u}r Bildung und Forschung (BMBF) under 
grant no.\ 05P15VTCA1 and M.S.\ by
the Institutional Strategy of the University of T\"{u}bingen (DFG, ZUK 63). 
The research of F.R.\ is supported by the US Department of Energy, Office of
Science under Contract No. DE-AC52-06NA25396 and by the DOE Early Career Program under Grant No. 2012LANL7033.

%%%%%%%%%%%%%%%%%%%%%%%%%%%

%
\end{document}